\documentclass[onwcolumn]{aastex631}
\usepackage{amsmath}
\usepackage{graphicx}

\usepackage {amsmath}
\usepackage{epsfig}

\graphicspath{{./}{figures/}}
\begin{document}
\title{Giga-Year Dynamical Evolution of Particles Around Mars}

\author{Yuying Liang}
\affiliation{Beihang University, Beijing, China}

\author{Ryuki Hyodo}
\affiliation{ISAS/JAXA, Sagamihara, Kanagawa, Japan}

\begin{abstract}

Particles of various sizes can exist around Mars. The orbits of large particles are mainly governed by Martian gravity, while those of small particles could be significantly affected by non-gravitational forces. Many of the previous studies of particle dynamics around Mars have focused on relatively small particles (radius of $r_{\rm p} \lesssim 100 \mu m$) for $\lesssim 10^{4}$ years. In this paper, using direct numerical orbital integration and analytical approaches, we consider Martian gravity, Martian $J_{2}$, the solar radiation pressure (SRP) and the Poynting-Robertson (PR) force to study the giga-year dynamical evolution of particles orbiting near the Martian equatorial plane with radius ranging from micrometer to meter. We also newly study the effect of the planetary shadow upon the particle dynamics. Our results show that small particles ($r_{\rm p} \lesssim 10 \mu m$) initially at $\lesssim 8$ Martian radii (below the orbit of today's Deimos) are quickly removed by the SRP due to eccentricity increase, resulting in a collision with Mars at the pericenter distnace. The orbits of larger particles ($r_{\rm p} > 10 \mu m$) slowly decay due to the PR forces (timescale of $> 10^{4}$ years). The planetary shadow reduces the sunlit area in the orbit and thus the efficiency of the PR drag force is reduced. However, we show that, even including the planetary shadow, particles up to $\sim 10$ cm in radius, initially at $\lesssim 8$ Martian radii, eventually spiral onto the Martian surface within $\sim 10^{9}$ years. Smaller particles require less time to reach Mars, and vice versa. Our results would be important to better understand and constrain the nature of the remaining particle around Mars in a context of giant impact hypothesis for the formation of Phobos and Deimos.

\end{abstract}


\section{Introduction}\label{sec:intro}

Various particles in size can be produced around Mars in different contexts. Micrometeoroidal impacts on Martian moons, Phobos and Deimos, can continuously create dust particles \citep{Ram13,Bra20}. Impact ejecta from Mars can hit Phobos and Deimos, producing particulate rings \citep{Ram17,Hyo19}. Alternatively, Phobos and Deimos may be formed as a byproduct of a giant impact \citep{Cra11,Ros16,Can18} and debris particles up to the order of meters in size \citep{Hyo17a,Hyo17b} could remain as a leftover of the moon-forming disk. Furthermore, a giant impact may produce an impact vapor around Mars, followed by condensation, forming small dust particles \citep{Hyo17a,Hyo17b,Hyo18}. A deeper understanding of the dynamical fate of such particles is, therefore, important to understand the nature of these physical processes as well as constrain the dynamical history of Mars-moon system.

The orbits of particles around Mars are governed by different forces, depending on their sizes. Although large particles are mainly controlled by the Martian gravity, the Martian oblateness and the non-gravitational forces such as  the solar radiation forces – e.g., the solar radiation pressure (SRP) and the Poynting-Robertson (PR) force – can significantly perturb the orbits of small particles.

The effect of the Martian oblateness moments on the particle dynamics around Mars was discussed in many literatures \citep[e.g.,][]{Kri96,Ham96} using the first zonal gravity coefficient, i.e., $J_{2}$ \citep{Mur99}. It does not cause a secular evolution of the semi-major axis and eccentricity, and the pericenter distance of a particle on average remain constant \citep{Kri96}. Higher-order zonal coefficients (e.g., $J_{4}$ and $J_{6}$) typically result in negligible secular change in the pericenter distance \citep{Mur99} because their values are much smaller than $J_{2}$; e.g., $J_{2}=1.96 \times 10^{-3}$ and $J_{4}=-1.54\times10^{-5}$ \citep{Zam15} for Mars. The $J_{6}$ term and higher-order terms are even smaller.

The SRP is a mechanical pressure exerted upon a particle surface due to the solar radiation. There is no secular effect of the SRP in the semi-major axis \citep{Kri96}. The PR force is caused by the nonuniform reemission of the sunlight that a particle absorbs \citep{Bur79,Mur99}, resulting in a decay in the particle orbit (i.e., semi-major axis) at a rate dependent on the particle size. The orbital decay becomes most significant when the particle size is comparable to the wavelength of the incident radiation (i.e., $\sim 1 \mu m$) \citep{Bur79,Mur99}.

Considering the above non-gravitational perturbation forces, depending on literature, many of the previous studies focused on the dynamical evolution of small dust particles (typically $r_{\rm p} \lesssim100 \mu m$) that are hypothetically released from Phobos and Deimos \citep{Kri96,Kri94,Ish96,Kri97,Sas99,Mak05,Liu20}. These studies reported that small particles of $r_{\rm p} \lesssim 1 \mu m$  are expected to collide onto Mars within a timescale of a few years or less, while larger particles ($r_{\rm p} \gtrsim 10 \mu m$) remain in orbit for $10^4$ years (and tens of years) at Deimos orbit (and Phobos orbit) \citep{Liu20}. 

However, a longer-term ($\gg 10^{4}$ years) dynamical evolution of larger particles ($r_{\rm p} >100 \mu m$) is not yet well studied. The magnitude of e.g., the PR force -- an important non-gravitational perturbation force for a long-term evolution -- depends on the strength of the solar radiation, which varies around a planet as a particle enters a shadowing area of a central planet. Such a planetary shadow effect may change a dynamical lifetime of particles around Mars as the planetary shadow effect was demonstrated at space debris around the Earth \citep{Hau12,Hau13}. \cite{Liu20} included the planetary shadow in their numerical model (see also \cite{Liu16}) for a dynamical timescale of $\sim 10^4$ years, although they did not study its effect.

In this paper, we focus on a giga-year dynamical evolution of particles around Mars ($\sim 10^9$ years) and establish the dynamical model by adding the PR force in addition to Martian $J_{2}$ and the SRP. We explicitly discuss the effect of the planetary shadow on particle dynamics. This paper is organized as follows. In Sec.~\ref{sec:dynamical}, the dynamical models concerning Martian $J_{2}$, the SRP, and the PR force are given and the mathematical criterion of the planetary shadow is shown. In Sec.~\ref{sec:result_wosun}, we study the case without the planetary shadow, and analytically discuss the long-term effect of each perturbation on the orbital elements using an averaging method and using numerical simulations. In Sec.~\ref{sec:with planetary shadow}, we include the planetary shadow effect. In Sec.~\ref{sec:diss}, we present a discussion. Finally, Section \ref{sec:summary} provides our conclusion.

\section{Dynamical Model for Particle Evolution} \label{sec:dynamical}

In this study, we assumed the situation where an impact on Mars occurred and debris particles were produced near the Martian equatorial plane \cite[e.g.,][see also Sec.~\ref{subsec:impact}]{Cra11}. Phobos and Deimos may be formed from impact debris (i.e., giant impact hypothesis), although the post-impact evolution and its likelihood are debated. Such a potential moon-forming impact on Mars is reported to distribute particles beyond the Martian synchronous orbit ($\sim 6$ Martian radii) with size ranging from micrometer to several meter \citep{Can18,Hyo17a,Hyo18}. In this study, we aim to understand the long-term dynamical fate of such debris particles, regardless of whether Phobos and Deimos are successfully formed or not; we do not include Phobos and Deimos.

We study dynamical evolution of a spherical particle (mass $m_{\mathrm{p}}$ and radius $r_{\mathrm{p}}$) released at an orbit around Mars. We consider that the particle's motion is governed by Martian gravity and non-gravitational perturbations related to the Sun, i.e., the solar radiation pressure (SRP) and the Poynting-Robertson (PR) force. We also consider the $J_{2}$ term of Mars. Mars is assumed to orbit around the Sun in a circular motion of period $T_{\mathrm{M}}$. We note that, for a particle around Mars (which is our target), the gravity of the Sun is not essential compared to other perturbation forces. In particular, up to $10 \mu m$ size, it is negligible to the SRP, e.g., its ratio to the SRP is $\sim 0.04$ ($\sim 0.6$) and is $\sim 0.08$ ($\sim 0.8$) to the gravity of Mars for a particle of $\sim \mu m$ ($\sim 10 \mu m$) at $1 a_{\mathrm{R}}$ and $3 a_{\mathrm{R}}$, where $a_{\mathrm{R}}$ is the Roche limit of Mars. Regardless of planetary shadow, according to \cite{Lor65}, the third-body gravity perturbations (e.g., the Sun's gravity in this paper) have zero secular effects in the semi-major axis of particles around the central body (e.g., Mars in this paper). In both cases with and without planetary shadow, the Sun's gravity perturbation would not influence the long-term evolution of the semi-major axis and would have little effect in the lifetime of particles (see Sec.~\ref{sec:result_wosun} and Sec.~\ref{sec:with planetary shadow}). Thus, the solar gravity is ignored in this study, although a more detailed study may be required.

\begin{figure}
    \centering
    \includegraphics[width=\textwidth]{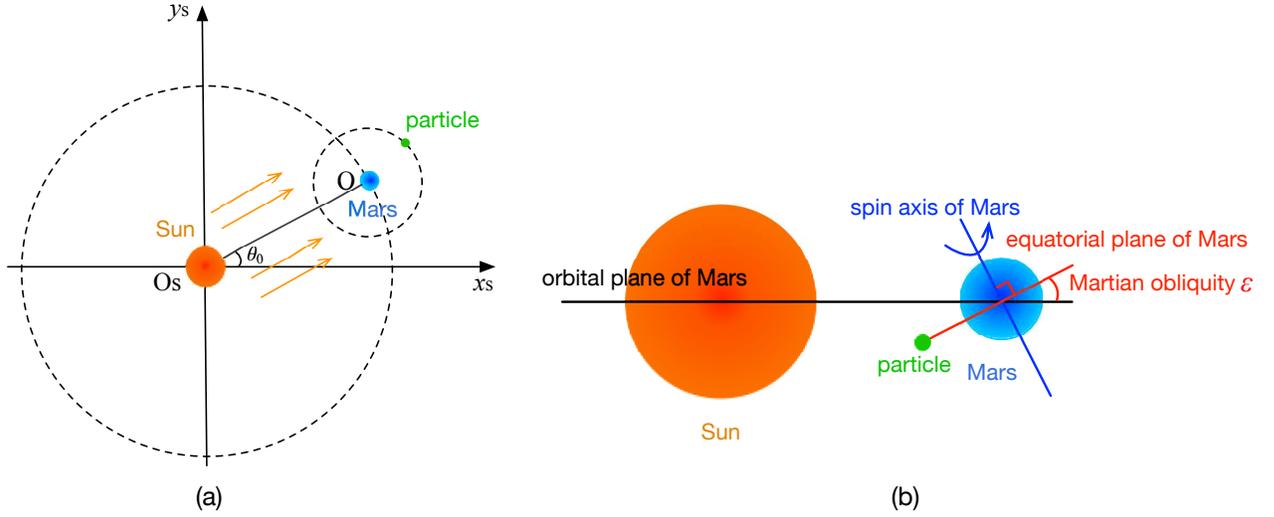}
    \caption{The schematic configuration of the Sun-Mars-particle system (panel (a)) and the obliquity of Mars (panel (b)). In panel (a), the $z_{\rm S}$-axis perpendicularly points outwards from the paper. Panel (b) presents the 3D configuration.}
\label{fig_SunMars_conf}
\end{figure}

\begin{figure}
    \centering
    \includegraphics[width=\textwidth]{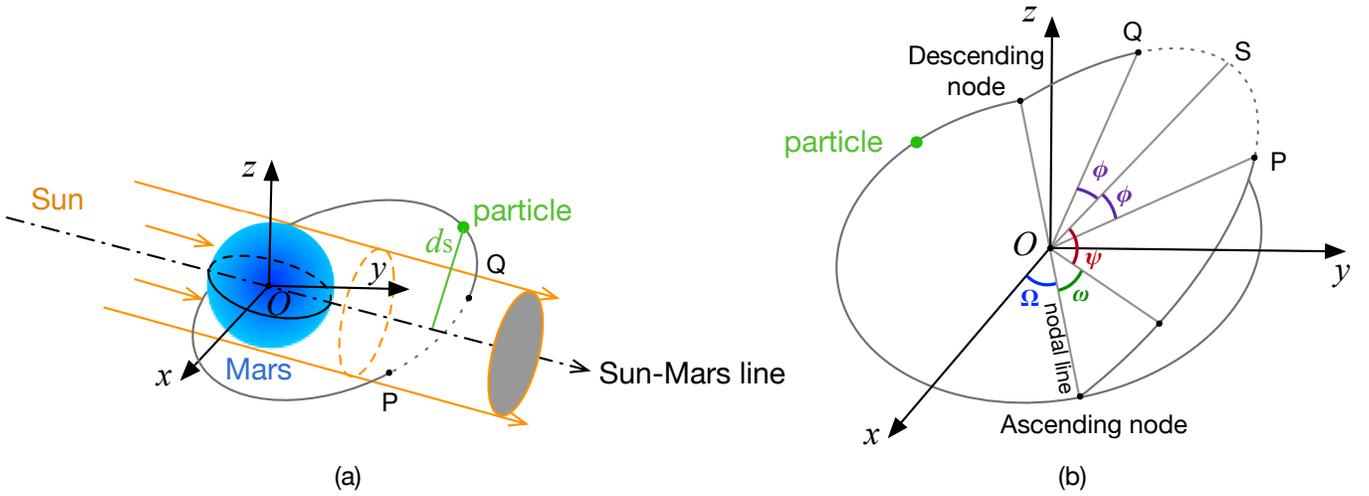}
    \caption{The schematic geometry of the planetary shadow (panel (a)) and its orientation (panel (b)). P and Q indicate the moment that a particle enters and exits the shadow area, respectively. S indicates the midpoint of arc PQ of the particle's orbit inside the planetary shadow cylinder, i.e., $\angle POS = \angle SOQ = \phi$ , where $2\phi$ indicates the shadow length. The OS direction is referred to as the orientation of the planetary shadow. The nodal line connects the ascending node and the origin of the Mars-centered inertial frame. $\omega+\psi$ indicates the angle between the orientation of the planetary shadow and the nodal line.}
\label{fig_shadow_conf}
\end{figure}

\subsection{Basic Governing Equations} \label{subsec:equations}

Here, to mathematically describe the motion of a particle, the Sun-centered inertial frame $\left(\mathrm{O}_{\mathrm{S}}-x_{\mathrm{S}}-y_{\mathrm{S}}-z_{\mathrm{S}}\right)$ and Mars-centered (O-$\it{x}$-$\it{y}$-$\it{z}$) inertial frame are established. The Sun is located at the origin and the $x_{\mathrm{S}}$-axis points to the Martian vernal equinox. The $z_{\mathrm{S}}$-axis is perpendicular to the Martian orbital plane outwards and the $y_{\mathrm{S}}$-axis follows the right-hand law. A Mars-centered inertial frame is defined in a similar manner but the origin is set as Mars and the $z$-axis is perpendicular to the Martian equatorial plane. Figure~\ref{fig_SunMars_conf} presents a schematic configuration of the Sun-Mars-particle system in the Sun-centered ineritial frame.

To describe the initial phase angle of Mars with respect to the Sun, a parameter $\theta_{0}$ is introduced, defined as the angle measured from the $x_{\mathrm{S}}$-axis to the Sun-Mars line, as shown in Fig.~\ref{fig_SunMars_conf}. The equations of motion for a particle around Mars are expressed in the Mars-centered inertial frame as \citep{Bur79, Kri96}
\begin{equation}
\left\{\begin{array}{l}
    \ddot{x}=-G M_{S} \frac{x}{|\boldsymbol{R}|^{3}}\left(1-J_{2} \psi_{2}\right)+B \frac{x+x_{S}}{\left|\boldsymbol{R}_{\mathrm{S}}\right|^{3}}+B \frac{1}{\left|\boldsymbol{R}_{\mathrm{S}}\right|^{2}}\left(-\frac{\dot{x}+\dot{x}_{\mathrm{S}}}{c}-\left(\frac{\boldsymbol{V}_{\mathrm{S}} \cdot \boldsymbol{R}_{\mathrm{S}}}{\left|\boldsymbol{R}_{\mathrm{S}}\right|^{2} \mathrm{c}}\right)\left(x+x_{\mathrm{S}}\right)\right) \\
    \ddot{y}=-G M_{S} \frac{y}{|\boldsymbol{R}|^{3}}\left(1-J_{2} \psi_{2}\right)+B \frac{y+y_{\mathrm{S}}\rm{cos}\epsilon}{\left|\boldsymbol{R}_{\mathrm{S}}\right|^{3}}+B \frac{1}{\left|\boldsymbol{R}_{\mathrm{S}}\right|^{2}}\left(-\frac{\dot{y}+\dot{y}_{\mathrm{S}}\rm{cos}\epsilon}{c}-\left(\frac{\boldsymbol{V}_{\mathrm{S}} \cdot \boldsymbol{R}_{\mathrm{S}}}{\left|\boldsymbol{R}_{\mathrm{S}}\right|^{2}}\right)\left(y+y_{\mathrm{S}}\rm{cos}\epsilon\right)\right) \\
    \ddot{z}=-G M_{S} \frac{z}{|\boldsymbol{R}|^{3}}\left(1-J_{2} \psi_{2}+J_{2} \Phi_{2}\right)+B \frac{z+y_{\mathrm{S}}\rm{sin}\epsilon}{\left|\boldsymbol{R}_{\mathrm{S}}\right|^{3}}+B \frac{1}{\left|\boldsymbol{R}_{\mathrm{S}}\right|^{2}}\left(-\frac{\dot{z}+\dot{y}_{\mathrm{S}}\rm{sin}\epsilon}{c}-\left(\frac{\boldsymbol{V}_{\mathrm{S}} \cdot \boldsymbol{R}_{\mathrm{S}}}{\left|\boldsymbol{R}_{\mathrm{S}}\right|^{2} \mathrm{c}}\right) (z+y_{\mathrm{S}}\rm{sin}\epsilon)\right),
\end{array}\right.
\label{eq_gov}
\end{equation}
where the first term on the right hand side is Martian gravity including the $J_{2}$ term. The second term on the right hand side indicates the SRP acceleration. The third term on the right hand side indicates the acceleration due to the PR force. $\boldsymbol{R}$ and $\boldsymbol{V}$ are the position and velocity vectors of a particle with respect to Mars. Their coordinates in the Mars-centered inertial frame are $\boldsymbol{R}=(x, y, z)$ and $\boldsymbol{V}=(\dot{x}, \dot{y}, \dot{z})$. $\boldsymbol{R}_{\rm S}$ and $\boldsymbol{V}_{\rm S}$ are the position and velocity vectors of Mars with respect to the Sun. Their coordinates in the Sun-centered inertial frame are $\boldsymbol{R}_{\rm S}=\left(x_{\mathrm{S}},y_{\mathrm{S}}, 0\right)$ and $\boldsymbol{V}_{\rm S}=\left(\dot{x}_{\mathrm{S}}, \dot{y}_{\mathrm{S}}, 0\right)$. $M_{\mathrm{S}}$ is the mass of Mars, $G$ is the gravitational constant, $c$ is the light speed, and $\epsilon$ is the obliquity (a tilt of the Martian spin axis to its orbital plane). The non-spherical gravity of Mars is truncated up to the $J_{2}$ term with $\psi_{2}$ and $\Phi_{2}$ expressed as \citep{Had00}

\begin{equation}
\left\{\begin{array}{l}
    \psi_{2}=\frac{R_{\mathrm{M}}^{2}}{|\boldsymbol{R}|^{2}}\left(\frac{15}{2}\left(\frac{z}{|\boldsymbol{R}|}\right)^{2}-\frac{3}{2}\right) \\
    \Phi_{2}=\frac{3 R_{\mathrm{M}}^{2}}{|\boldsymbol{R}|^{2}}
\end{array},\right.
\label{eq_phi}
\end{equation}
where $R_{\mathrm{M}}$ denotes the average radius of Mars. The second and third terms on the right hand side are characterized by a coefficient $B$ as \citep{Rub13,Bur79}
\begin{equation}
    B=\frac{\pi Q r_{\mathrm{p}}^{2} F_{\mathrm{M}} a_{\mathrm{M}}^{2}}{ c M_{\mathrm{p}}},
\label{eq_B}
\end{equation}
where $F_{\mathrm{M}}$ is the insolation at Mars distance from the Sun and $a_{\mathrm{M}}$ is the average Sun-Mars distance. $Q$ is the radiation pressure efficiency for the particle and is assumed to be 1 for simplicity. $M_{\mathrm{p}}$ is the particle's mass.

The orbit of a particle with respect to Mars can be characterized by Kepler elements $(a, e, i, \omega, \Omega, f)$, where $a$ is the semi-major axis and $i$ is the inclination from the Martian equatorial plane. The current value of the Martian obliquity is $\epsilon \sim 25^{\circ}$. Table \ref{tab_parameter} lists the values of the parameters used in this study (referring to NASA's Mars Fact Sheet)

\begin{table}[]
    \centering
    \caption{Parameters for the Sun-Mars system used in this paper}
    \begin{tabular}{c c c c}
    \hline
   {G} & {$T_{\rm M}$}& {$m_{\rm S}$} & {$F_{\rm M}$}  \\
    \hline
    $6.67\times 10^{-11} \mathrm{N\cdot m^2/kg^2}$ & 686.98 days &   $6.42\times 10^{23} \rm kg $  &    $5.86\times 10^{2} \rm W/m_{2} $ \\
   \hline
   {$a_{\rm M}$} & {$R_{\rm M}$}& {$c$} & {$J_{2}$}  \\
   \hline
   $2.28\times 10^{11} \rm m$ & $3.39\times 10^{6} \rm m$  &  $3.00\times 10^{8} \rm m/s $  &    $1.96\times 10^{-3}$ \\
   \hline
    \end{tabular}
\label{tab_parameter}
\end{table}

\subsection{planetary shadow} \label{subsec:planetary shadow}

In this study, Mars orbits around the Sun, while particles orbit around Mars. Thus, a particle around Mars may enter the planetary shadow area during the orbit, depending on the obliquity $\epsilon$. The planetary shadow may play a non-negligible effect in the dynamics of a particle around a planet \citep{Rub13}. Here, to analytically demonstrate the effects of the planetary shadow, we employ the basic umbra assumption that the Sun is assumed to be far enough from Mars and the solar rays are considered to be parallel. Figure \ref{fig_shadow_conf}(a) presents the geometric configuration of such cylindrical shadow and its orientation is illustrated in Fig.~\ref{fig_shadow_conf}(b). As presented in Fig.~\ref{fig_shadow_conf}(b), $\psi$ denotes the orientation of the midpoint of the passage arc of particle's orbit inside the planetary shadow and $2 \phi$ denotes the length of the planetary shadow. $\psi-\phi$ and $\psi+\phi$ correspond to the shadow entrance and exit (i.e., $\mathrm{P}$ and $\mathrm{Q}$ moment in Fig.~\ref{fig_shadow_conf}(b)).

Due to the aberration of starlight, the orientation of the shadow rotates about the Sun-Mars lie by an angle $\alpha$ \citep{Rub13} and $\alpha \approx \tan \alpha=\frac{\left|\boldsymbol{V}_{\mathrm{S}}\right|}{c}$. $\alpha$ is small, e.g., $\sim 10^{-5}$ for Mars. Thus, we ignore this angle in this study. \cite{Rub13} discussed the effect of this angle for particles on a planar and an inclined circular orbit. They qualitatively demonstrated that there are no significant terms of the order of $\frac{\left|\boldsymbol{V}_{\mathrm{S}}\right|}{c}$ for a planar circular orbit of a particle. Considering this angle, \cite{Rub13} quantitatively showed a null result to the order of $\frac{\left|\boldsymbol{V}_{\mathrm{S}}\right|}{c}$ for an inclined circular particle orbit. However, there is an error in the mathematical derivation at Eq. (20) of \cite{Rub13} and we present it in Sec.~\ref{subsec:Rubincam}.

To examine whether a particle is inside the planetary shadow of Mars, we introduce an auxiliary frame as ($\mathrm{O} -x^{\prime}-y^{\prime}-z^{\prime}$). The $z^{\prime}$-axis is perpendicular to the Martian orbital plane ($x^{\prime}-y^{\prime}$), while the $z$-axis is perpendicular to the Martian equatorial plane ($x-y$). This frame is only employed for the planetary shadow examination. Based on the assumption of cylindrical shadow, if a particle around Mars is inside the planetary shadow of Mars, its position coordinates in the ($\mathrm{O} -x^{\prime}-y^{\prime}-z^{\prime}$) frame must obey the following constraint:
\begin{equation}
    d_{S} \cdot \cos \left(\sin ^{-1}\left(\frac{z^{\prime}}{d_{S}}\right)\right) \leq R_{\mathrm{M}} \cdot \cos \left(\sin ^{-1}\left(\frac{z^{\prime}}{R_{\mathrm{M}}}\right)\right),
\label{eq_ds}
\end{equation}
where $d_{S}$ denotes the normal distance from the position of the particle to the Sun-Mars line and $d_{S} \cdot \cos \left(\sin ^{-1}\left(\frac{z^{\prime}}{d_{S}}\right)\right)$ denotes the length of its projection onto the Martian orbital plane. $R_{\mathrm{M}} \cdot \cos \left(\sin ^{-1}\left(\frac{z^{\prime}}{R_{\mathrm{M}}}\right)\right)$ denotes the radius of the spherical small circle intersected by the horizontal plane parallel to the Martian orbital plane at particle's height $z^{\prime}$ and it determines the boundary of the planetary shadow at this horizontal plane. Based on the exact coordinate of the particle, Eq.~(\ref{eq_ds}) is used to determine whether a particle is inside the shadow cylinder (and the length of the planetary shadow) regardless of particle's orbital distance and obliquity. We note that Eq.~(\ref{eq_ds}) is mathematically the same as those used in \cite{Liu16}, although their expressions look different.

Furthermore, based on the semi-major axis alone, we propose a quick evaluation on the length of the planetary shadow for particles moving in a planar orbit ($i=0$) around Mars as follows. For an arbitrary value of $\epsilon$, the particle's orbit is partially inside the shadow cylinder, depending on its orbital distance (i.e., semi-major axis $a$); if the semi-major axis is large enough, the particle's orbit is entirely out of the shadow cylinder, while if the semi-major axis is small enough, it is partially inside the shadow cylinder. For a given $\epsilon$, there is a critical orbital distance at which a particle's orbit barely intersects the shadow cylinder and below which the particle passes through the shadow cylinder in one orbit. This critical orbital distance $a_{\text {intersect}}$ is $R_{\mathrm{M}} / \sin \epsilon$. A similar critical distance is also obtained by \cite{Rub13}. 

When $\epsilon=0$, a particle passes through the cylinder shadow in one orbit regardless of the particle's distance. In particular, if $a<a_{\rm intersect}$, the portion of the whole orbit under the planetary shadow has a length of an arc of $2 \sin ^{-1}\left(R_{\mathrm{M}} / a\right)$ in radian (i.e., $\phi=\sin ^{-1}\left(R_{\mathrm{M}} / a\right)$). For Phobos orbit ($\sim 3$ Martian radius) and Deimos orbit ($\sim 7$ Martian radius), about $12 \%$ and about $5 \%$ of the whole orbit are under the planetary shadow, respectively. As the orbital radius of a particle becomes smaller, the portion of the planetary shadow becomes larger and correspondingly the effect of the planetary shadow would become stronger. Thus, the planetary shadow would exert an important effect on the particle dynamics when the non-gravitational effects are included (see Sec.~\ref{subsec:compar wo sun} and Sec.~\ref{subsec:depend obliquity}).

\subsection{Numerical Settings} \label{subsec:numerical setting}

We numerically solved the three-dimensional motion of a particle around Mars under Martian $J_{2}$, the SRP, and the PR force. The $J_{3}$ term was neglected in this study because it would have a negligible influence on the secular evolution of particle dynamics \citep{Liu20}. We used a fourth-order Hermite scheme with variable step size \citep{Kok04}. The initial position of Mars was set at $\left(-a_{\mathrm{M}}, 0,0\right)$ in the Sun-centered inertial frame. 
The initial conditions of particles around Mars were described by Kepler orbital elements (i.e., $a_{0}, e_{0}, i_{0}, \omega_{0}, \Omega_{0}, f_{0}$). As a reasonable simplification, the angular variables at $t=0$, i.e., $\omega_{0}$ and $\Omega_{0}$, were set to zero. These angular variables are quickly randomized compared to the dynamical timescale interested here (e.g., giga-year) and thus an arbitrary initial choice does not affect our results.

The initial semi-major axis of a particle $a_{0}$ was set as follows: we set $a_{0}=a_{\mathrm{R}}$ for particles that were assumed to be at Phobos orbit, where $a_{\mathrm{R}}$ is the Roche limit of Mars (about $a_{\mathrm{R}} \sim 9116 \mathrm{~km}$); we set $a_{0}=3 a_{\mathrm{R}}$ for particles at Deimos orbit. Dependence on the eccentricity was studied with $e_{0}=0-0.7$. We set $i_{0}=0$ because we were interested in particles produced by an impact near the equatorial plane. We leave the initial inclination dependence for future study. We study dependence on particle size ($r_{\rm p} \sim 0.1 \mu$m $-10$m; \cite{Hyo17a}). We set a particle's bulk density as $3 \mathrm{~g}^{\cdot} \mathrm{cm}^{-3}$ for simplicity \citep{Hyo18,Pig18}.

To include the effects of the planetary shadow, both the SRP and PR terms in Eq.~(\ref{eq_gov}) were switched to zero when the particle enters the planetary shadow area and recovered when the particle exits the shadow area. The `light-off' and `light-on' judgement was mathematically determined by the planetary shadow criterion (i.e., Eq.~(\ref{eq_ds})). 

The numerical simulations were obtained by discretely integrating the continuous dynamical flow with an acceptably small step size. However, such implementation unavoidably results in some error in the determination of entrance and exit moments for planetary shadow. To judge the planetary shadow as accurately as possible, a forward and backward integration method based on dichotomy was adopted until the smallest step size was achieved. The smallest step size used in the integration was $1 \mathrm{~m}$ ($\sim 10^{-7} a_{\mathrm{R}}$); we confirmed the convergence of our numerical results.

The numerical simulation was terminated when either of the following conditions were met: 1) the particle collides with the Martian surface; 2) the particle travels beyond the Hill radius of Mars, which would indicate a gravitational escape of the particle.

\section{Results without planetary shadow}
\label{sec:result_wosun}

In this section, we show the results of numerical simulations that include Martian $J_{2}$, the SRP, the PR. Here, the planetary shadow is not included. We step-by-step demonstrate their effects on the particle dynamical evolution around Mars. Section \ref{subsec:J2&SRP} shows the averaging effect of Martian $J_{2}$ and the SRP. Section \ref{subsec:simu for J2&SRP&PR} presents the numerical results that include Martian $J_{2}$, the SRP, and the PR.

\subsection{Martian $J_{2}$ and SRP} \label{subsec:J2&SRP}

Here, we briefly review the dynamical aspects of a particle orbital evolution regarding the $J_{2}$ term and the SRP. The averaged disturbing potential of Martian $J_{2}$ term $\bar{R}_{J_{2}}$ is expressed as \citep{Lan15}
\begin{equation}
    \bar{R}_{J_{2}}=\frac{G m_{\mathrm{S}} R_{\mathrm{M}}^{2} J_{2}}{2 a^{3}\left(1-e^{2}\right)^{3 / 2}}\left(1-\frac{3}{2} \sin ^{2} i\right),
\label{eq_J2}    
\end{equation}
which yields that $\left< \frac{\mathrm{d} a}{\mathrm{d} t}\right>_{J_{2}}=\left< \frac{\mathrm{d} e}{\mathrm{d} t} \right>_{J_{2}}=0$ (Substituting Eq.~(\ref{eq_J2}) to Eqs.~(\ref{eq_Lagrange_a}) and (\ref{eq_Lagrange_e})), indicating that Martian $J_{2}$ does not cause a secular change in the semi-major axis and eccentricity. Thus, Martian $J_{2}$ alone does not produce a secular change in the energy and the shape of the particle's orbit around Mars, and the particle does not spiral into Mars by Martian $J_{2}$ \citep{Kri96}.

The averaged disturbing potential of the SRP can be expressed as \citep{Kri96}
\begin{equation}
\begin{aligned}
    \bar{R}_{\rm SRP} = \frac{3}{2} \sigma n^{2} a^{2} e & \left[\cos \omega\left(\cos \Omega \cos \lambda_{S} + \sin \Omega \sin \lambda_{S} \cos \epsilon\right) \right. \\ 
    & \left. +\sin \omega\left(-\sin \Omega \cos i \cos \lambda_{S}+\cos \Omega \cos i \sin \lambda_{\mathrm{S}} \cos \epsilon\right) + \sin i \sin \lambda_{S} \sin \epsilon \right],
\label{eq_SRP}
\end{aligned}
\end{equation}
where $\sigma$ stands for the ratio of the SRP force to the solar gravity force and $\lambda_{\mathrm{S}}$ stands for the longitude of the Sun measured in the Martian orbital plane from the $x$-axis. According to Eq.~(\ref{eq_SRP}), the following equations can be derived (Substituting Eq.~(\ref{eq_SRP}) to Eqs.~(\ref{eq_Lagrange_a}) and (\ref{eq_Lagrange_e})):
\begin{equation}
    \left<\frac{\mathrm{d} a}{\mathrm{d} t}\right>_{\rm SRP}=0
\label{eq_a_SRP}
\end{equation}
and \citep[see also][]{Ish96,Lan15}
\begin{equation}
\begin{aligned}
    \left<\frac{\mathrm{d} e}{\mathrm{d} t}\right>_{\rm SRP} = -\frac{3}{2} n \sigma \sqrt{1-e^{2}} & \left[-\sin \omega\left(\cos \Omega \cos \lambda_{\mathrm{S}} + \sin \Omega \sin \lambda_{\mathrm{S}} \cos \epsilon\right) \right. \\ 
    & \left. +\cos \omega \left(-\sin \Omega \cos i \cos \lambda_{\mathrm{S}}+\cos \Omega \operatorname{cosi} \sin \lambda_{\mathrm{S}} \cos \epsilon\right)\right] .
\label{eq_e_SRP}
\end{aligned}
\end{equation}
Thus, the SRP perturbation exerts a secular change on the eccentricity but not on the semi-major axis. The maximum value in one oscillation of eccentricity, denoted by $e_{\max}$, is dominated by the strength of the radiation pressure that is related to the particle size \citep{Kri96,Sas99}.

In short conclusion, Martian $J_{2}$ alone does not change the energy and the shape of the particle's orbit around Mars, and collision with Mars is not driven by Martian $J_{2}$ alone. With the SRP added, the two-body energy of the particle's orbit still remains unchanged (Eq.~(\ref{eq_a_SRP})), but the shape of its orbit alters in each period of the particle's motion (Eq.~(\ref{eq_e_SRP})). The particle is then forced to deviate from its initial orbit due to the oscillation of eccentricity. As the eccentricity increases towards $e_{\max}$, the pericenter distance may become smaller than the Martian radius, potentially resulting in a collision with Mars. This process can be predicted by the analytical evolution of eccentricity, especially the dependence of $e_{\max}$ on the particle size \citep{Kri96,Sas99}. The increasing eccentricity in one oscillation is the first factor responsible for the decreasing value of the pericenter distance and thereby for the collision with Mars. For example, at Phobos orbits, particle sizes smaller than $\sim 40 \mu m$ collide with Mars when $e_{\max}$ is increased to $0.64$ or larger \citep{Kri96}, while at Deimos orbit, the critical size of the particles that reaches the Martian surface at their pericenter is $\sim 20 \mu m$, as demonstrated by Fig.~9 of \cite{Kri96} and Fig.~1 of \cite{Sas99}. Our numerical code reproduced their results \cite{Liu20} also confirmed that, for these small particles at the orbits around Phobos, the SRP is the most important perturbation force compared to both Martian $J_{2}$ and $J_{3}$.

\subsection{Simulations under Martian $J_2$, SRP and PR without planetary shadow} 
\label{subsec:simu for J2&SRP&PR}

Here, using our own numerical simulations, we revisit the long-term evolution of a particle orbiting at Phobos or Deimos orbit considering Martian $J_{2}$, the SRP, and the PR (here, without planetary shadow). Section \ref{subsec:general} presents the general results. Section \ref{subsec:depend size} and Section \ref{subsec:depend eccen} demonstrate the dependence on particles sizes and initial eccentricities, respectively.

\subsubsection{General results}
\label{subsec:general}

Following \cite{Rub13} and \cite{Bur79}, the PR force is decomposed to 6 terms concerning the relative particle's motion with respect to Mars and the motion of Mars around the Sun as follows (see Appendix \ref{sec:PR}),
\begin{equation}
    \boldsymbol{F}_{\rm PR}=\frac{B}{|\boldsymbol{d}|^{2}}\left[-\frac{\boldsymbol{V}_{\mathrm{S}}}{c}-\frac{\boldsymbol{V}_{\mathrm{S}} \cdot \boldsymbol{R}_{\mathrm{S}}}{c|\boldsymbol{d}|^{2}} \boldsymbol{d}-\frac{\boldsymbol{V}_{\mathrm{S}} \cdot \boldsymbol{R}}{c|\boldsymbol{d}|^{2}} \boldsymbol{d}-\frac{\boldsymbol{V}}{c}-\frac{\boldsymbol{V} \cdot \boldsymbol{R}_{\mathrm{S}}}{c|\boldsymbol{d}|^{2}} \boldsymbol{d}-\frac{\boldsymbol{V} \cdot \boldsymbol{R}}{c|\boldsymbol{d}|^{2}} \boldsymbol{d}\right]=\sum_{i=1}^{6} \boldsymbol{F}_{{\rm PR}, i},
\label{eq_FPR}
\end{equation}
where $\boldsymbol{F}_{{\rm PR}, i}(i=1,2, \ldots, 6)$ denotes each term in $\boldsymbol{F}_{\rm PR}$ from left to right. $\boldsymbol{d}$ is its position vector with respect to the Sun. Thus, the secular perturbation on $a$ due to each $\boldsymbol{F}_{{\rm PR}, i}$ is shown as follows \citep[][see Appendix \ref{sec:PR}]{Rub13}:

For $\boldsymbol{F}_{\mathrm{PR}, 1}, \boldsymbol{F}_{\mathrm{PR}, 2}$, and $\boldsymbol{F}_{\mathrm{PR}, 6}$
\begin{equation}
    \left<\frac{\mathrm{d} a}{\mathrm{d} t}\right>_{F_{\mathrm{PR}, 1}}=\left<\frac{\mathrm{d} a}{\mathrm{d} t}\right>_{F_{\mathrm{PR}, 2}}=\left<\frac{\mathrm{d} a}{\mathrm{d} t}\right>_{F_{\mathrm{PR}, 6}}=0 .
\label{eq_a_PR1&2&6}
\end{equation}
The non-zero effect on the semi-major axis due to the PR force is characterized by $\boldsymbol{F}_{\mathrm{PR}, 3}, \boldsymbol{F}_{\mathrm{PR}, 4}$, and $\boldsymbol{F}_{\mathrm{PR}, 5}$. Assuming a small obliquity (i.e., $\cos^{2} \epsilon \sim 1$ and $\sin^{2} \epsilon \sim 0$; see Appendix \ref{sec:PR}), these terms in total can be expressed as
\begin{equation}
    \left<\frac{\mathrm{d} a}{\mathrm{d} t}\right>_{F_{\mathrm{PR}}}^{\epsilon \sim 0} \approx-\frac{1}{n} \frac{B\left|\boldsymbol{V}_{\mathrm{S}}\right|\left|\boldsymbol{R}_{\mathrm{S}}\right||\boldsymbol{R}|}{c|\boldsymbol{d}|^{4}} \cos i-\frac{2}{n} \frac{B|\boldsymbol{V}|}{c|\boldsymbol{d}|^{2}}\left(1+\frac{1}{4}\left(1+\cos ^{2} i\right)\right) .
\label{eq_a_PR}
\end{equation}
Hence, the cumulative PR force exerts a long-term decay of the semi-major axis on the particles around Mars, which was also reported in \cite{Rub13}.

\begin{figure}
    \centering
    \includegraphics[width=\textwidth]{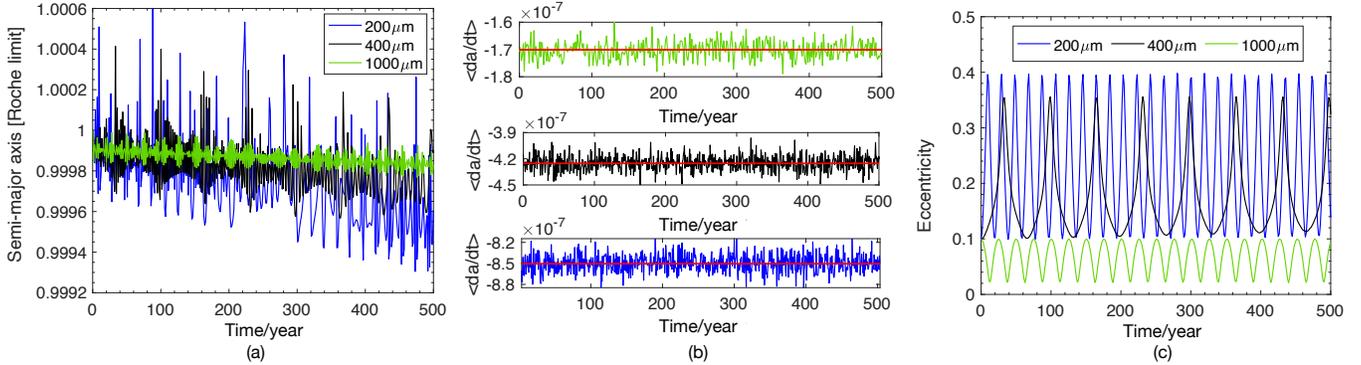}
    \caption{Evolution of $a$ (panel (a)), $\left<da/dt\right>$ (panel (b)), and $e$ (panel (c)) under Martian $J_{2}$, the SRP, and the PR force without the planetary shadow. Here, the particles sized are $r_{\rm p}=200 \mu m$ (blue), $400 \mu m$  (black), and $1000 \mu m$ (green). The red lines in panel (b) indicate the analytical result at $t=0$ for all tested particle sizes. The unit of the vertical axis ($\left<da/dt\right>$) in panel (b) is the Roche limit per year. Here, $a_{\rm R} \sim 9000$ km.}
\label{fig_dadt}
\end{figure}

Figure \ref{fig_dadt} presents the results of numerical simulations under the PR force for particles with $r_{\mathrm{p}}=200 \mu m$ (blue lines), $400 \mu m$ (black lines) and $1000 \mu m$ (green lines) released at $a_{0}=1 a_{\mathrm{R}}$ and $e_{0}=0.1$. Figure \ref{fig_dadt}(a) shows the short-term evolution of the semi-major axis (for $\sim 500$ years). A decrease in the semi-major axis is observed. A similar decay phenomenon in the semi-major axis was also observed by \cite{Mak05} for $15 \mu m$ and $7.5 \mu m$ sized particles at Deimos orbit where the same dynamical model was used.

Figure \ref{fig_dadt}(b) shows the decay rate of the semi-major axis for particles of $r_{\mathrm{p}}=200 \mu m$ (blue lines), $400 \mu m$ (black lines) and $1000 \mu m$ (green lines), respectively. The analytical approximations of $\left<\frac{\mathrm{d} a}{\mathrm{d} t}\right>_{F_{\mathrm{PR}}}^{\epsilon \sim 0}$ (Eq.~(\ref{eq_a_PR}); red lines in Fig.~\ref{fig_dadt}(b)) are also shown using the values at $t=0 $; $\left<\frac{\mathrm{d} a}{\mathrm{d} t}\right>_{F_{\mathrm{PR}}}^{\epsilon \sim 0} \approx-8.505 \times 10^{-7} a_{\mathrm{R}}$ per year, $-4.252 \times 10^{-7} a_{\mathrm{R}}$ per year, and $-1.701 \times 10^{-7} a_{\mathrm{R}}$ per year for $r_{\mathrm{p}}=200,400,1000 \mu m$, respectively. Figure \ref{fig_dadt}(b) demonstrates that the numerical results of the decay rate of the semi-major axis are in agreement with the analytical approximations. This indicates that the analytical approximation of $\left<\frac{\mathrm{d} a}{\mathrm{d} t}\right>_{F_{\mathrm{PR}}}^{\epsilon \sim 0}$ (Eq.~(\ref{eq_a_PR})) can describe the decay rate of the semi-major axis and its long-term evolution. Figure \ref{fig_dadt}(b) shows that a smaller particle decays more quickly. This is because the magnitude of the PR force is $\left|\boldsymbol{F}_{\rm PR}\right| \propto \frac{1}{r_{\mathrm{P}}}$ (Eq.~(\ref{eq_FPR}) and Eq.~(\ref{eq_a_PR})).

Figure \ref{fig_dadt}(c) presents the evolution of eccentricity for particles of $r_{\mathrm{p}}=200 \mu m$ (blue lines), $400 \mu m$ (black lines) and $1000 \mu m$ (green lines). Under Martian $J_{2}$ and the SRP, the evolution of eccentricity shows an oscillating pattern that can be described by its period and amplitude \citep{Kri96,Sas99}. Adding the PR force, Fig.~\ref{fig_dadt}(c) shows that such oscillation still exists and the amplitude and period do not significantly change. Our numerical results of the full dynamics agree with those obtained from the integration with the averaged eccentricity in \cite{Mak05}, where only the variation caused by the decay of the averaged semi-major axis is accounted. \cite{Mak05} reported that the decay of the semi-major axis due to $\mathrm{PR}$ does not affect the oscillations of the eccentricity; $e_{\max}$ and period of oscillation are mainly governed by Martian $J_{2}$ and the SRP \citep{Kri96}.

In short, adding the PR force cumulatively exerts a decay in the semi-major axis of particles around Mars, which is the key difference from the effects of only Martian $J_{2}$ and the SRP. The long-term decrease in the semi-major axis is the second factor responsible for the decrease in the pericenter distance and can be approximated by the analytical estimation (Eq.~(\ref{eq_a_PR})).

\begin{figure}
    \centering
    \includegraphics[width=\textwidth]{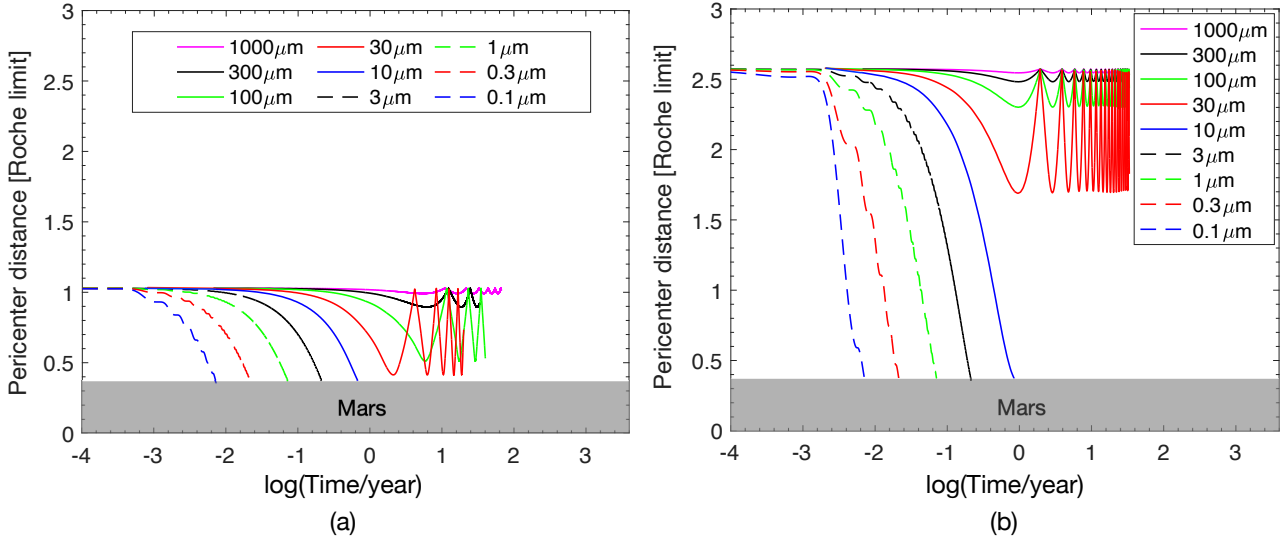}
    \caption{Evolution of pericenter distance under Martian $J_{2}$, the SRP, and the PR for the particles at Phobos orbit (panel (a)) and Deimos orbit (panel (b)). Particles sizes are depicted by different colors and lines as shown in the figure legend. The grey bar indicates Mars. Here, $a_{\rm R} \sim 9000$ km.}
\label{fig_allsize}
\end{figure}

\subsubsection{Dependence on particle size}
\label{subsec:depend size}

Here, we focus on the evolution of the pericenter distance for different particle sizes with the PR force considered. We set $a_{0}=r_{\rm Phobos}$ or $r_{\rm Deimos}$ with $e_{0}=0$. $r_{\mathrm{p}}=0.1,0.3,10,30,100,300$, and $1000 \mu m$.

Figure \ref{fig_allsize} presents the evolution of the pericenter distance of particles starting at Phobos orbit (Fig.~\ref{fig_allsize}(a)) and at Deimos orbit (Fig.~\ref{fig_allsize}(b)). Figure \ref{fig_allsize} shows that particles whose $r_{\mathrm{p}}$ is smaller than $\sim 10 \mu m$ collide with Mars within $\sim 1$ year in both cases of Phobos and Deimos orbital distances \cite[see also][]{Kri96,Liu20}. This can be explained by the fact that the magnitude of the PR force is much smaller than the SRP force according to their definitions in Eq.~(\ref{eq_gov}). Thus, the evolution of the particle's motion in a short period (e.g., $\sim 1$ year) is dominated by the SRP, and can be estimated by the analytical approximations derived by \cite{Kri96} without the PR force.

According to \cite{Kri96}, the value of $e_{\max}$ in the eccentricity evolution under Martian $J_{2}$ and the SRP can be analytically approximated by $e_{\max} \sim 14 / r_{\mathrm{p}}$ for particles at Deimos orbit, where $r_{\mathrm{p}}$ indicates the particle size in micrometer. This indicates that the critical size of particles that hits the Martian surface at their pericenter is $\sim 20 \mu m$ for those at Deimos orbit. Later, \cite{Sas99} numerically updated that the critical size of particles at Deimos orbit is $\sim 7 \mu m$. In Phobos case, $e_{\max} \sim 54 / r_{\mathrm{p}}$ was analytically reported, which is applicable for particles of $r_{\mathrm{p}}>100 \mu m$, while the numerical simulations indicated particles of $r_{\mathrm{p}}<40 \mu m$ collide with Mars. Thus, it can be concluded that, considering the PR force, the critical size of particles whose $e_{\max}$ is large enough to cause rapid collision on Mars is similar to those under Martian $J_{2}$ and the SRP (without the PR force), i.e., $\sim 40 \mu m$ at Phobos orbit and $\sim 20 \mu m$ at Deimos orbit. These critical sizes at Phobos and Deimos orbits are of the same order as our numerical results $(\sim 10 \mu m)$.

For larger particles $\left(r_{\mathrm{p}} \gtrsim 10 \mu m\right)$, their pericenter distances in both Fig.~\ref{fig_allsize}(a) and Fig.~\ref{fig_allsize}(b) oscillate above the Martian radius for $\sim 100$ years. This reconfirmed \cite{Mak05} that particles of $r_{\mathrm{p}} \gtrsim 10 \mu m$ at Deimos orbit can stay orbiting around Mars for more than hundreds of years. 

\begin{figure}
    \centering
    \includegraphics[width=\textwidth]{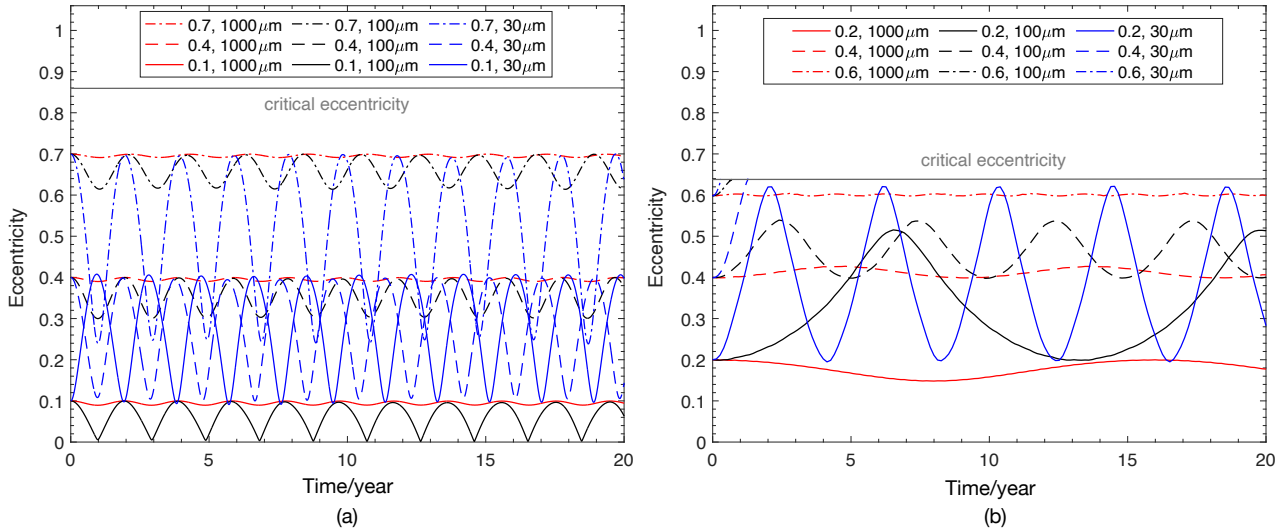}
    \caption{Eccentricity evolution of the particles at Deimos orbit (panel (a)) and Phobos orbit (panel (b)) under Martian $J_{2}$, the SRP, and the PR with non-zero $e_{\rm 0}$. The grey lines indicate $e=0.86$ (the critical value for particles at Deimos orbit) in panel (a) and $e=0.64$ (the critical value for particles at Phobos orbit) in panel (b).}
\label{fig_ecc}
\end{figure}

\subsubsection{Dependence on initial eccentricity}
\label{subsec:depend eccen}

Besides the semi-major axis, the pericenter distance, $a(1-e)$, depends on eccentricity. This part is devoted to the question whether a larger initial value of eccentricity results in a quicker drop of pericenter distance below the Martian radius for particles with $r_{\mathrm{p}} \gtrsim 10 \mu m$. For this purpose, we set $i_{0}=0$ and $r_{\mathrm{p}}=30$ and $100 \mu m$. For particles at Deimos orbit, we set $a_{0}=r_{\text {Deimos}}$, and $e_{0}=0.1,0.4$, and $0.7$; for particles at Phobos orbit, we set $a_{0}=r_{\text {Phobos}}$, and $e_{0}=$ 0.2, 0.4, and 0.6. According to Fig.~\ref{fig_allsize}(b), with $e_{0}=0$, these particle sizes do not lead to a rapid collision with Mars. The total simulation time is 503 years but only the evolution of the first 20 years is presented here (Fig.~\ref{fig_ecc}).

Figure \ref{fig_ecc} presents the evolution of eccentricity of particles starting at Deimos orbit (Fig.~\ref{fig_ecc}(a)) and at Phobos orbit (Fig.~\ref{fig_ecc}(b)) under Martian $J_{2}$, the SRP, and the PR forces. The critical eccentricity at Phobos orbit that reaches the Martian surface at the pericenter is $e \sim 0.64$ (grey line in Fig.~\ref{fig_ecc}(b)) and the one at Deimos orbit is $e \sim 0.86$ (grey line in Fig.~\ref{fig_ecc}(a)). The collision is regarded to happen when the value of the eccentricity reaches the critical eccentricity (ignoring the variation in the semi-major axis). To plot Fig.~\ref{fig_ecc}, the maximum value of initial eccentricity that we take is $0.7$, which directly reduces the initial pericenter distance by about $2 / 3$ of that in the zero-eccentricity case.

For particles at Deimos orbit, all tested sizes with $e_{0}=0.7$ (dot-dashed lines) shows that the oscillation of their eccentricities always stay below their initial values $e_{0}$, which makes the pericenter distance in an oscillation larger than the initial value and thereby the particles do not reach the Martian surface. The outcomes with $e_{0}=0.4$ (dashed lines) and $e_{0}=0.1$ (solid lines) are similar for most of the tested particle sizes, except that at $e_{0}=0.1$, the oscillating eccentricity of the particle of $30 \mu \mathrm{m}$ (blue solid line) is pumped up above $e_{0}$, resulting in a drop of pericenter distance, whereas no collision occurs. According to \cite{Kri96}, the value of $e_{\max}$ is approximated by $54 / r_{\mathrm{p}}$ and $14 / r_{\mathrm{p}}$ for particles at Phobos orbit and Deimos orbit, respectively, so the SRP force exerted upon a smaller particle is stronger $\left(e_{\max} \sim 14 / 30 \mu m \sim 0.47\right)$ and the variation of eccentricity oscillation becomes larger. Thus, the simulation results for particles of $30 \mu m$ (blue lines) shows the largest variation in eccentricity oscillation. As demonstrated by Fig.\ref{fig_ecc}(a), the tested sizes (30 $\mu \mathrm{m}, 100$ $\mu m$, and $1000 \mu m$) initially released on an elliptic orbit do not hit Mars within the simulation time (503 years). Together with the results shown in Fig.~\ref{fig_allsize}(b), particles with $r_{\mathrm{p}} \gtrsim 10 \mu m$ starting at Deimos orbit do not collide onto Mars within hundreds of years.

For particles at Phobos orbit, the situation varies greatly. According to Fig.~\ref{fig_allsize}(a), the particles at Phobos orbit with $r_{\mathrm{p}} \gtrsim 10 \mu m$ and $e_{0}=0$ do not collide with Mars within hundreds of years. After increasing their initial eccentricity, according to Fig.~\ref{fig_ecc}(b), some of these particles still remain orbiting Mars (without rapid collision) within hundreds of years. For example, for the particles of $1000 \mu m$ with $e_{0}=0.2$ (red solid line), $e_{0}=0.4$ (red dashed line), and $e_{0}=0.6$ (red dotdashed line), and those of $100 \mu m$ with $e_{0}=0.2$ (black solid line) and $e_{0}=0.4$ (black dashed line), their eccentricity oscillations in a period of 503 years stay below the critical eccentricity at Phobos orbit $\left(e \sim 0.64 \right)$.

However, the particles of $30 \mu m$ collide with Mars rapidly within few years as their initial eccentricities are increased from 0 (red solid line in Fig.~\ref{fig_allsize}(a)) to $0.4$ (blue dashed line in Fig.~\ref{fig_ecc}(b)) and $0.6$ (blue dot-dashed line in Fig.~\ref{fig_ecc}(b)). With $e_{0}=0, r_{\mathrm{p}}=30 \mu m$ case does not result in a collision with Mars (Fig.~\ref{fig_allsize}(a)). For a short timescale of few years, the PR force, as a cumulative effect, has little effect on the semi-major axis. Here, the SRP is the dominant perturbation that is responsible for the collision caused by increasing the eccentricity of the particle to the critical eccentricity ($e \sim 0.64)$ and decreasing its pericenter distance to the Martian radius. According to Fig.~\ref{fig_ecc}(b), collision happens $\left(r_{\mathrm{p}}=30 \mu m\right.$ with $\left.e_{0}>0.4\right)$ before half oscillation of eccentricity. 

Increasing initial eccentricities $(0.4$ and $0.6)$ for particles of $30 \mu m$ can be explained by \cite{Ham96}, where they plotted phase portraits to describe the dynamical evolution of particles from Phobos. In the phase portraits, each trajectory indicates an eccentricity evolution starting from a specific initial eccentricity for a specific particle. The polar coordinates of each point on each trajectory indicates the value of oscillating eccentricity and the phase angle of the Sun at a specific epoch. Here, we take particles of $30 \mu m$ as an example. The last phase portrait of Fig.~7 of \cite{Ham96} describes the eccentricity evolution of $30 \mu m$ particles at Phobos orbit and each trajectory on the plot indicates an eccentricity evolution starting from a specific $e_{0}$. As we can be observe from their Fig.~7, the eccentricity increases along the trajectory starting from the origin (i.e., $e_{0}=0$) to its maximum value of about $0.6$, corresponding to the minimum pericenter distance of $0.4 a_{\mathrm{R}}$, which is not small enough for a collision. The maximum eccentricity then increases to $0.65$, corresponding to a pericenter distance of $0.35 a_{\mathrm{R}}$ as the location of the $e_{0}$ moves rightwards along the horizontal axis to $0.4$. According to \cite{Ham96} for small particles (e.g., $r_{\mathrm{p}}=30 \mu m$) at Phobos orbit, the critical initial eccentricity that avoids a collision is $e_{0} \sim 0.1-0.4$, which agrees with our results and is specified to a value between $0.2$ and $0.4$ by Fig.~\ref{fig_ecc}(b).

Therefore, it can be concluded from our results that the increasing initial eccentricity up to $e_{0}=0.7$ does not lead to collisions with Mars at Deimos orbit. In Phobos case, increasing $e_{0}$ from 0 up to $0.6$, particles of $r_{\mathrm{p}}=10^{3} \mu m$ still survive without new collisions occurred. These particles would orbit around Mars until their pericenter distances decreases via the cumulative effect of the PR force to reach the Martian surface. The potential collision time is not significantly changed compared with the cases of $e_{0}=0$ in Section \ref{subsec:depend size}. Particles of $r_{\mathrm{p}}<100 \mu m$ at Phobos orbit with $e_{0} \gtrsim 0.4$ collide with Mars within few years, while these particles with $e_{0}=0.1$, during such a short timescale, orbit around Mars without collision.

\section{Results with planetary shadow}
\label{sec:with planetary shadow}

Inclusion of the planetary shadow effect at the non-gravitational forces could influence the motion of the particle slowly but the accumulating perturbation in a long timescale may not be negligible \citep{Kri96,All62}. Indeed, for a near-Earth orbit, \cite{All62} showed that the value of $a \dot{e}$ under SRP is increased by $25 \%$ at maximum $(a=8000 \mathrm{~km}$ and $e=0.1)$ when the Earth's shadow is considered.

Including the planetary shadow, the relative change in the semi-major axis under SRP is within 2-3\% for particles around Mars \citep{Kri96}. However, they did not consider the shadowed PR force and their integration time was limited to tens of years. The long-term cumulative effect of the shadowed PR force is, therefore, still not well understood, which motivated our study. Here, we focus on the particle dynamics around Mars under the effects of the planetary shadow from the analytical and numerical aspects. In particular, we first provide an analytical upper bound of the pericenter distance of a particle in Sec.~\ref{subsec:pmax}; then numerical simulations considering the planetary shadow are presented in Sec.~\ref{subsec:simu with sun}; a comparison with the case without the planetary shadow is shown in Sec.~\ref{subsec:compar wo sun}; next, we discuss the dependence on obliquity in Sec.~\ref{subsec:depend obliquity}; a comparison to the analytical arguments of \cite{Rub13} is presented in Sec.~\ref{subsec:Rubincam}.

\subsection{Upper bound of pericenter distance $\left(p_{\max}\right)$}
\label{subsec:pmax}

The planetary shadow only alters the secular evolution of the semi-major axis due to the solar radiation, i.e., SRP and PR. The general description of the semi-major axis evolution that include the planetary shadow can be expressed as
\begin{equation}
    \left<\frac{\mathrm{d} a}{\mathrm{d} t}\right>_{\rm shadow}=\frac{1}{2 \pi} \frac{2}{n}\left(\int_{0}^{2 \pi} \boldsymbol{F}(f) \cdot \hat{\boldsymbol{t}} \mathrm{d} f-\int_{\psi-\phi}^{\psi+\phi} \boldsymbol{F}(f) \cdot \hat{\boldsymbol{t}} \mathrm{d} f\right) ,
\label{eq_a_sunshadow_total}
\end{equation}
where $\boldsymbol{F}=\boldsymbol{F}_{\rm SRP}+\boldsymbol{F}_{\rm PR}$. As presented in Fig.~\ref{fig_shadow_conf}(b), $\psi$ denotes the orientation of the midpoint of the passage arc of particle's orbit inside the planetary shadow and $2 \phi$ denotes the length of the planetary shadow \citep{Rub13}. $\psi-\phi$ and $\psi+\phi$ correspond to the shadow entrance and exit (i.e., $\mathrm{P}$ and $\mathrm{Q}$ moment in Fig.~\ref{fig_shadow_conf}(b)). Equation (\ref{eq_a_sunshadow_total}) can be rewritten as
\begin{equation}
    \left<\frac{\mathrm{d} a}{\mathrm{d} t}\right>_{\rm shadow}=\frac{1}{2 \pi} \frac{2}{n}\left(\int_{0}^{2 \pi} \boldsymbol{F}_{\rm PR}(f) \cdot \hat{\boldsymbol{t}} \mathrm{d} f-\int_{\psi-\phi}^{\psi+\phi} \boldsymbol{F}_{\rm PR}(f) \cdot \hat{\boldsymbol{t}} \mathrm{d} f+\int_{\psi+\phi}^{\psi-\phi} \boldsymbol{F}_{\rm SRP}(f) \cdot \hat{\boldsymbol{t}} \mathrm{d} f\right) .
\label{eq_a_sunshadow_expand}
\end{equation}
Here, $\int_{0}^{2 \pi} \boldsymbol{F}_{\rm SRP}(f) \cdot \hat{\boldsymbol{t}} \mathrm{d} f = 0$.

The analytical discussion regarding the secular perturbation of the shadowed SRP on the semi-major axis is discussed in detail in Appendix \ref{sec:SRP and J2}. In summary, the shadowed SRP alone triggers a long-term periodic variation in the semi-major axis, which is first discovered by \cite{Hau13}. Such a variation is flattened when Martian $J_{2}$ is added. Thus, the overall effects of the shadowed SRP and Martian $J_{2}$ on the semi-major axis are ignored in the analytical arguments and we only discuss the shadowed PR force as follows.

Equation (\ref{eq_a_sunshadow_expand}) is simplified as follows,
\begin{equation}
    \left<\frac{\mathrm{d} a}{\mathrm{d} t}\right>_{\rm shadow} \simeq \frac{1}{2 \pi} \frac{2}{n}\left[\int_{0}^{2 \pi} \boldsymbol{F}_{\rm PR}(f) \cdot \hat{\boldsymbol{t}} \mathrm{d} f-\int_{\psi-\phi}^{\psi+\phi} \boldsymbol{F}_{\rm PR}(f) \cdot \hat{\boldsymbol{t}} \mathrm{d} f\right] \equiv \left<\frac{\mathrm{d} a}{\mathrm{d} t}\right>_{F_{\mathrm{PR}}}-\left<\frac{\mathrm{d} a}{\mathrm{d} t}\right>_{F_{\mathrm{PR, shadow}}}.
\label{eq_a_sunshadow_PR}
\end{equation}
The first term in Eq.~(\ref{eq_a_sunshadow_PR}) describes the effect of the PR force without the planetary shadow, which is analytically discussed in Sec.~\ref{subsec:general} and Appendix ~\ref{sec:PR}. The second term of Eq.~(\ref{eq_a_sunshadow_PR}) indicates the decay rate of the semi-major axis in the shadow area due to each component $\boldsymbol{F}_{{\rm PR}, i}$, given as
\begin{equation}
    \left<\frac{\mathrm{d} a}{\mathrm{d} t}\right>_{\boldsymbol{F}_{\mathrm{PR}, i, \rm shadow}} \equiv \frac{1}{2 \pi} \frac{2}{n} \int_{\psi-\phi}^{\psi+\phi} \boldsymbol{F}_{\mathrm{PR}, i}(f) \cdot \hat{\boldsymbol{t}} \mathrm{d} f,
\label{eq_a_inshadow_PR}
\end{equation}
where $i=1,2, \ldots, 6$. Most of the components cause little effect on the integral of Eq.~(\ref{eq_a_inshadow_PR}) and they can be negligible (see Eqs.~(\ref{eq_PR1_integ})-(\ref{eq_PR6_final}) in Appendix \ref{sec:PR}). Only the shadowed $\boldsymbol{F}_{{\rm PR}, 4}$ and $\boldsymbol{F}_{{\rm PR}, 5}$ are dominant factors. The expressions of the integrals in Eq.~(\ref{eq_a_inshadow_PR}) are as follows (see Eqs.~(\ref{eq_PR4_integ})-(\ref{eq_PR5_final}) in Appendix \ref{sec:PR}),
\begin{equation}
\begin{aligned}
    \left<\frac{\mathrm{d} a}{\mathrm{d} t}\right>_{F_{{\mathrm{PR}}, 4, \rm shadow}} &= -\frac{1}{2 \pi} \frac{B}{c|\boldsymbol{d}|^{2}} \frac{2}{n} \int_{\psi-\phi}^{\psi+\phi} \boldsymbol{V} \cdot \hat{\boldsymbol{t}} \mathrm{d} f \\ 
    & = -\frac{1}{2 \pi} \frac{2}{n} \frac{B|\boldsymbol{V}|}{c|\boldsymbol{d}|^{2}}(2 \phi+2 \sin 2 \Omega \cos (2 \omega+2 \psi) \sin 2 \phi)
\label{eq_a_inshadow_PR4}
\end{aligned}
\end{equation}
and
\begin{equation}
\begin{aligned}
    \left<\frac{\mathrm{d} a}{\mathrm{d} t}\right>_{F_{{\mathrm{PR}}, 5, \rm shadow}} &= -\frac{1}{2 \pi} \frac{B}{c|\boldsymbol{d}|^{2}} \frac{2}{n} \int_{\psi-\phi}^{\psi+\phi} \boldsymbol{V} \cdot \boldsymbol{R}_{\mathrm{S}} \cdot \boldsymbol{d} \cdot \hat{\boldsymbol{t}} \mathrm{d} f \\ 
    & = -\frac{1}{2 \pi} \frac{2}{n} \frac{B|\boldsymbol{V}|}{c|\boldsymbol{d}|^{2}}\left(\frac{1}{2}\left(1+\cos ^{2} \epsilon\right) \phi+\frac{1}{2}\left(1-\cos ^{2} \epsilon\right) \sin (2 \Omega+2 \omega+2 \psi) \sin 2 \phi\right) .
\label{eq_a_inshadow_PR5}
\end{aligned}
\end{equation}
The parameter $2 \phi$ can be approximated as the one of a planar circular orbit sheltered from Mars and $\phi=\arcsin \left(R_{\mathrm{M}} / a\right)$. To further simplify Eq.~(\ref{eq_a_inshadow_PR4}) and Eq.~(\ref{eq_a_inshadow_PR5}) and to capture essential physical dependence, $\psi$, $\Omega$, and $\omega$ are hoped to be rewritten, although \cite{Hau13} showed that $\psi$ cannot be explicitly expressed by $\Omega$ and $\omega$.

The minimum of Eq.~(\ref{eq_a_inshadow_PR4}) and Eq.~(\ref{eq_a_inshadow_PR5}) (i.e., the minimum of $\left<\frac{\mathrm{d} a}{\mathrm{d} t}\right>_{F_{{\mathrm{PR}}, \rm shadow}}$) can be given analytically as (see Eq.~(\ref{eq_PR_inshadow}) and Eq.~(\ref{eq_PR_inshadow_minimum}) in Appendix \ref{sec:sun_shadow})
\begin{equation}
\begin{aligned}
    \left<\frac{\mathrm{d} a}{\mathrm{d} t}\right>_{F_{{\mathrm{PR}}, \rm shadow}} &= \left<\frac{\mathrm{d} a}{\mathrm{d} t}\right>_{F_{{\mathrm{PR}}, 4, \rm shadow}}  + \left<\frac{\mathrm{d} a}{\mathrm{d} t}\right>_{{F_{{\mathrm{PR}}, 5, \rm shadow}}}  \\
    & \geq -\frac{1}{2 \pi} \frac{2}{n} \frac{B|\boldsymbol{V}|}{c|\boldsymbol{d}|^{2}}\left(\frac{1}{2}\left(1+\cos ^{2} \epsilon\right) \phi+2 \phi+\frac{1}{2}\left(5-\cos ^{2} \epsilon\right) \sin 2 \phi\right).
\label{eq_a_inshadow_total}
\end{aligned}
\end{equation}
The first and second terms in Eq.~(\ref{eq_a_inshadow_total}) are proportional to the shadow length parameter $2 \phi$ and they share the common term $\frac{2}{n} \frac{B|\boldsymbol{V}|}{c|\boldsymbol{d}|^{2}}$ with Eq.~(\ref{eq_a_PR})  i.e., $\left<\frac{\mathrm{d} a}{\mathrm{d} t}\right>_{F_{\mathrm{PR}}}$. Thus, they indicate that the negative cumulative effect of the PR force is compensated by the positive effect of the planetary shadow. The overall effects of both the PR force and the planetary shadow are determined by the shadow length. Substituting Eq.~(\ref{eq_a_inshadow_total}) to Eq.~(\ref{eq_a_sunshadow_PR}) yields the lowest decay rate, $\left<\frac{\mathrm{d} a}{\mathrm{d} t}\right>_{\rm shadow,lowest}$, as
\begin{equation}
    \left<\frac{\mathrm{d} a}{\mathrm{d} t}\right>_{\rm shadow} = \left<\frac{\mathrm{d} a}{\mathrm{d} t}\right>_{F_{\mathrm{PR}}}-\left<\frac{\mathrm{d} a}{\mathrm{d} t}\right>_{F_{\mathrm{PR}, \rm shadow}} \leq \left<\frac{\mathrm{d} a}{\mathrm{d} t}\right>_{\rm shadow,lowest},
\label{eq_a_sunshadow_inequality}
\end{equation}
where
\begin{equation}
    \left<\frac{\mathrm{d} a}{\mathrm{d} t}\right>_{\rm shadow,lowest} = -\frac{2}{n} \frac{B|\boldsymbol{V}|}{c|\boldsymbol{d}|^{2}}\left[1+\frac{1}{4}\left(1+\cos ^{2} i\right)-\frac{1}{2 \pi}\left(\frac{1}{2}\left(1+\cos^{2} \epsilon\right) \phi + 2 \phi + \frac{1}{2} \left(5-\cos ^{2} \epsilon \right) \sin 2 \phi \right) \right] <0 .
\label{eq_a_sunshadow_lowest}
\end{equation}

Equation (\ref{eq_a_sunshadow_lowest}) is a negative value, indicating a decay of the semi-major axis. The terms on the right hand side of Eq.~(\ref{eq_a_sunshadow_lowest}), i.e., $\frac{1}{2}\left(1+\cos ^{2} \epsilon\right) \phi+2 \phi+\frac{1}{2}\left(5-\cos ^{2} \epsilon\right) \sin 2 \phi$, becomes maximum when $\cos \epsilon=1$, i.e., the value of obliquity $\epsilon$ is small, (e.g., for current Martian obliquity), indicating the lowest decay rate of the semi-major axis. Given the small obliquity, we can also assume that the particle passes through the shadow cylinder in one orbit at Phobos or Deimos distance.

Now, using Eq.~(\ref{eq_a_sunshadow_lowest}), we can analytically give the lowest decay rate of the semi-major axis and correspondingly the upper bound of the semi-major axis $a_{\max}$ at any epoch. The upper bound of the pericenter distance $p_{\max}$ can be, then, constrained by the upper bound of the semi-major axis via $p_{\max}=a_{\max}$. This is because the range of the pericenter distance $p$ is guaranteed by $0 \leq p \leq a_{\max}$, i.e., the semi-major axis must be larger than $p_{\max}$ or they are equal only if $e=0$. 

In this study, we aim to estimate the upper limit of the dynamical lifetime of particles under the PR drag using $p_{\max}$ (this process is detailed in Appendix \ref{sec:upper}). Below, using $p_{\max}$ (that is, with an assumption of $e=0$), our main focus is the long-term dynamical evolution of inwardly drifting particles until a collision with Mars occurs. In reality, particles would be dragged by the PR force more efficiently than those estimated by the lowest decay rate (i.e., by Eq.~(\ref{eq_a_sunshadow_lowest})). Furthermore, the real value of eccentricity is $e > 0$ for particles orbiting around Mars. For a larger value of eccentricity, the real pericenter distance becomes smaller than $p_{\max}$ where $e=0$ is assumed.

\begin{figure}
    \centering
    \includegraphics[width=0.7\textwidth]{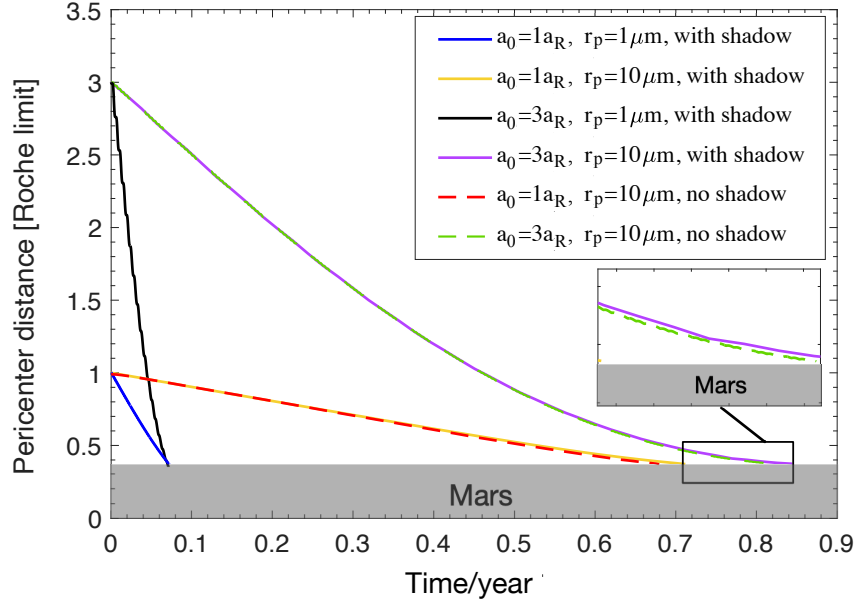}
    \caption{Evolution of the pericenter distance under Martian $J_{2}$, the SRP, the PR. Solid and dashed lines indicate the cases with and without the planetary shadow, respectively. Here, particles are released from $a_{\rm 0}=1 a_{\rm R}$ and $3 a_{\rm R}$ with $r_{\rm p}=1 \mu m$ and $10 \mu m$. The grey bar indicates Mars. Here, $a_{\rm R} \sim 9000$ km.}
\label{fig_smallsize}
\end{figure}

\begin{figure}
    \centering
    \includegraphics[width=0.7\textwidth]{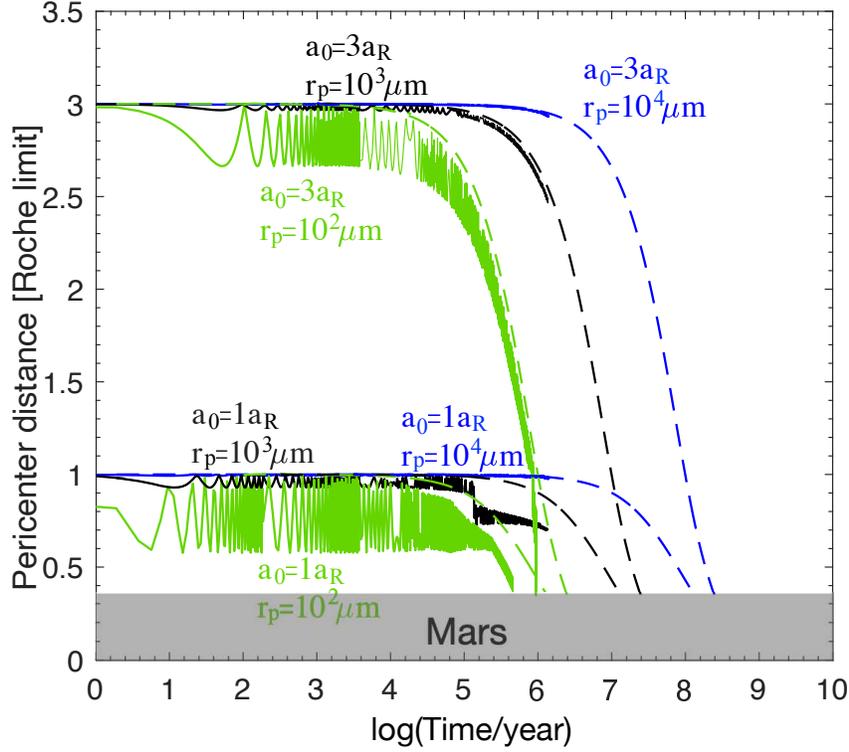}
    \caption{Comparison between analytical $p_{\rm max}$ (dashed lines; Eq.~(\ref{eq_a_sunshadow_lowest})) and numerical simulations (solid lines). The particles are released from $a_{\rm 0}=1 a_{\rm R}$ and $3 a_{\rm R}$ under Martian $J_{2}$, the SRP, the PR, and the planetary shadow with $r_{\rm p}=10^2 \mu m$ (green), $10^3 \mu m$ (black), and $10^4 \mu m$ (blue). The grey bar indicates Mars. Here, $a_{\rm R} \sim 9000$ km.}
\label{fig_compare}
\end{figure}

\begin{figure}
    \centering
    \includegraphics[width=0.7\textwidth]{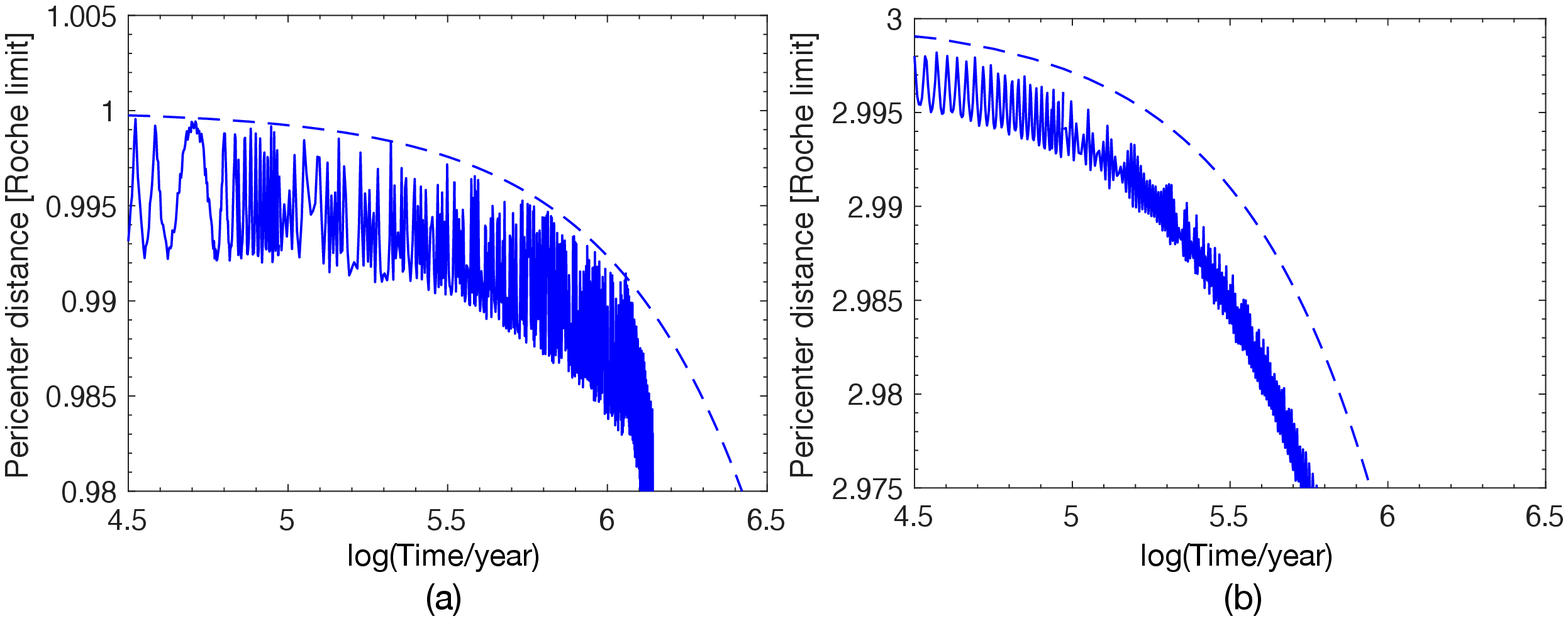}
    \caption{Same as Fig.~\ref{fig_compare} but a zoom-in view. Here, $r_{\rm p}=10^4 \mu m$ with $a_{\rm 0}=1a_{\rm R}$ (panel (a)) and $r_{\rm p}=10^4 \mu m$ with $a_{\rm 0}=3a_{\rm R}$ (panel (b)).}
\label{fig_zoom}
\end{figure}

\subsection{Simulations with planetary shadow}
\label{subsec:simu with sun}

Here, we conduct the following numerical simulations: $a_{0}=1 a_{\mathrm{R}}$ and $3 a_{\mathrm{R}}, e_{0}=0$, and $r_{\mathrm{p}}=1$, $10$, $10^{2}$, $10^{3}$, and $10^{4} \mu m$. Figure \ref{fig_smallsize} shows the effects of the planetary shadow on the evolution of the pericenter distance for particles released from $a_{0}=1 a_{\mathrm{R}}$ and $3 a_{\mathrm{R}}$ with $r_{\mathrm{p}}=1 \mu m$ and $10 \mu m$. As presented in Fig.~\ref{fig_smallsize}, for small particles $\left(r_{\mathrm{p}} \lesssim 10 \mu m\right)$, the evolution of the pericenter distance is barely changed when we include the planetary shadow (blue and black lines), compared with those without the planetary shadow (red and green lines). Small particles fall onto Mars within 1 year with/without the planetary shadow. Therefore, similar to the PR force, the cumulative effect of the planetary shadow can be ignored for small particles with $r_{\mathrm{p}} \lesssim 10 \mu m$.

Figure \ref{fig_compare} shows the evolutions of the pericenter distances for large particles ($r_{\rm p}=10^{2}$, $10^{3}$, and $10^{4} \mu m$). The dashed lines indicate $p_{\max}$ derived in Sec.~\ref{subsec:pmax} (Eq.~(\ref{eq_a_sunshadow_lowest}); see also Appendix \ref{sec:upper}) and the solid lines indicate the results of our numerical simulations. Due to the limited computation capability, the evolution up to $\sim 10^{9}$ years cannot be fully numerically performed, and the longest simulation time is limited to $\sim 10^{6}-10^{7}$ years depending on cases. 

According to the numerical results (solid lines), the pericenter distances of all sizes here show a decrease within the current simulation time. Particles of $r_{\rm p}=10^{2} \mu m$ released at $1 a_{\mathrm{R}}$ and $3 a_{\mathrm{R}}$ (green solid lines) fall onto Mars after $\sim 5 \times 10^{5}$ years and $\sim 10^{6}$ years, respectively. The case of $r_{\rm p}=10^{3} \mu m$ released at $1 a_{\mathrm{R}}$ (black solid line) shows a sudden decrease at $\sim 5.2 \times 10^{5}$ years due to an increase in the eccentricity. \cite{Mak05} reported the similar phenomenon when they simulated the eccentricity evolution of particles of 15$\mu m$, 10$\mu m$, and 7.5$\mu m$ from Deimos under Martian $J_{2}$, the SRP, and the PR force. Such phenomenon can be explained by the phase portraits of the dynamical system by \cite{Ham96} (cf. Fig. 3 and Fig. 7 of \cite{Ham96} for particles from Deimos and Phobos, respectively). Based on the phase portraits, \cite{Mak05} provided the detailed explanations and related numerical simulations together with a comparison to other particles sizes. In short, there exists bifurcations on the phase portraits (eccentricity vs solar phase angle) of specific particle sizes, e.g., particles of $1000 \mu m$ from Phobos in this paper and those of $7.5 \mu m$ from Deimos of \cite{Mak05}. As the semi-major axis gradually decreases by the PR force, the new eccentricity trajectory bifurcates from the initial one at a specific energy level and the sudden increase in the eccentricity occurs. 

$p_{\max}$(dashed lines; assuming $e=0$) always stays larger than the numerical simulations (solid lines; $e>0$). To clarify the details of Fig.~\ref{fig_compare}, Figure \ref{fig_zoom} presents a zoom-in view of the results of $r_{\rm p}=10^{4} \mu m$ with $a_{0}=1 a_{\mathrm{R}}$ (blue solid line in the left panel) and $3 a_{\mathrm{R}}$ (blue solid line in the right panel). Figures \ref{fig_compare} and \ref{fig_zoom} shows that the solid lines stay below the corresponding dashed lines (Eq.~(\ref{eq_a_sunshadow_lowest})), suggesting that the numerical results of the pericenter distance $(e>0)$ are constrained by their $p_{\max}$ (with $e=0$). Thus, the upper limit of the dynamical lifetime of the pericenter distance up to $\sim 10^{9}$ years can be also predicted according to its $p_{\max}$.

As presented by the dashed lines of Fig.~\ref{fig_compare}, $p_{\max}$ for all tested particles $\left(r_{\mathrm{p}} \lesssim 10^{4} \mu m\right.$ and $\left.a_{0} \lesssim 3 a_{\rm R}\right)$ drops to the Martian radius within $\sim 10^{9}$ years, indicating that collisions with Mars happen for all tested particles within $\sim 10^{9}$ years. It is important to note that the estimations using $p_{\max}$ assumed $e=0$, while in reality $e>0$, that is, a collision with Mars can happen earlier than the predictions by $p_{\max}$. Therefore, particles of $r_{\mathrm{p}} \lesssim 10^{4} \mu m$ are likely to collide with Mars with less than the age of the solar system.

\subsection{Comparison between cases with and without the planetary shadow}
\label{subsec:compar wo sun}

Here, using analytical arguments, we discuss the lifetime of particles having $r_{\mathrm{p}} \gtrsim 10 \mu m$ without the planetary shadow and compare with those including the planetary shadow. In the case without the planetary shadow, the decay rate of the semi-major axis (also $p_{\max}$) by the PR force is approximated by Eq.~(\ref{eq_a_PR}) (see also Appendix \ref{sec:PR}). When we include the planetary shadow, the decay rate of the pericenter distances is described by Eq.~(\ref{eq_a_sunshadow_lowest}).

\begin{figure}
    \centering
    \includegraphics[width=0.7\textwidth]{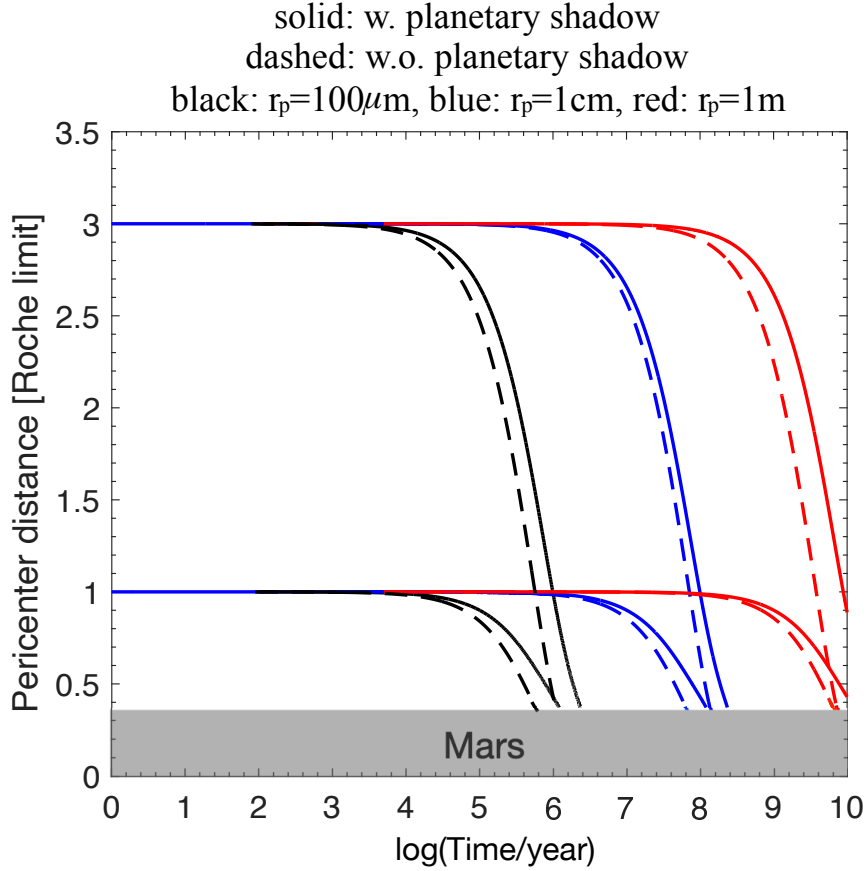}
    \caption{Comparison between the pericenter distances with (solid lines; $p_{\rm max}$ of Eq.~(\ref{eq_a_sunshadow_lowest})) and without (dashed lines; Eq.~(\ref{eq_a_PR})) the planetary shadow. Black, blue, and red colors indicate $r_{\rm p}=100$ $\mu$m, $1$ cm, and $1$ m, respectively. Here, $a_{\rm R} \sim 9000$ km.}
\label{fig_shadow}
\end{figure}

Figure \ref{fig_shadow} presents the analytical solutions of the pericenter distance with and without the planetary shadow (solid and dashed lines, respectively). We study particles of $100 \mu m$ (black), $1 \mathrm{~cm}$ (black), and $1 \mathrm{~m}$ (red). To be consistent with Fig.~\ref{fig_compare}, particles are assumed to start from both $3 a_{\mathrm{R}}$ (at Deimos orbit) and $1 a_{\mathrm{R}}$ (at Phobos orbit). In all cases without the planetary shadow (dashed lines), the pericenter distances of the particles decrease to the Martian radius by the PR force within $10^{10}$ years. The decay rate including the planetary shadow (solid line) is lower and thus it takes longer time to reach the Martian radius. Even including the planetary shadow, particles with $r_{\mathrm{p}} \lesssim 10^{4} \mu m$ (i.e., $r_{\mathrm{p}} \lesssim 1 \mathrm{~cm}$; black and blue solid lines) reach the Martian radius by the PR force within $10^{9}$ years.

\begin{figure}
    \centering
    \includegraphics[width=0.9\textwidth]{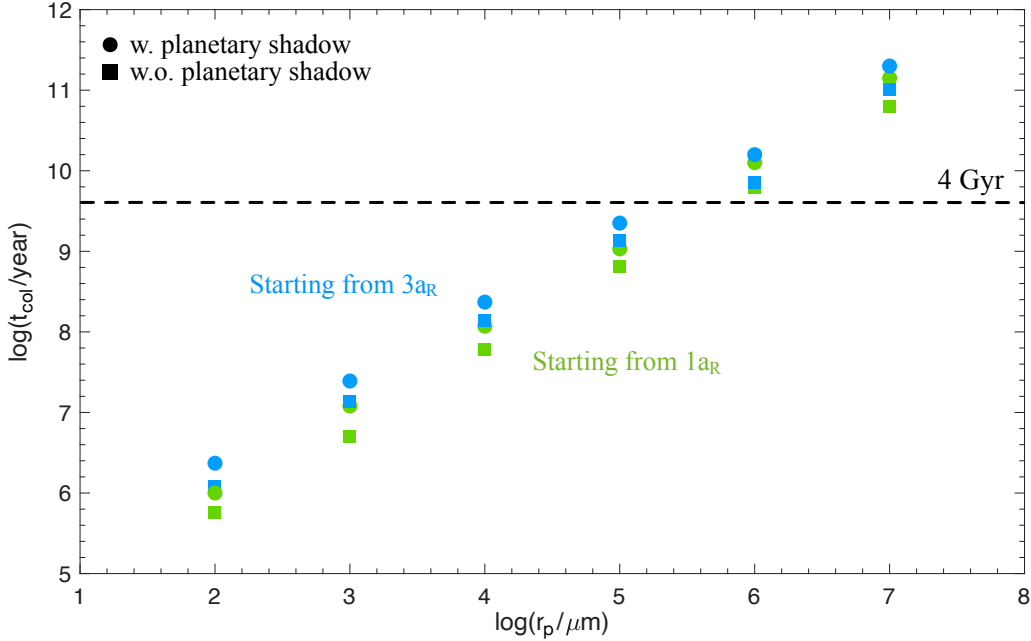}
    \caption{Time to collide for particles starting from $1 a_{\rm R}$ (green) and $3 a_{\rm R}$ (blue) with (circles) and without (squares) the planetary shadow. The black dashed line indicates $4$ Gyr. The particles are regarded to collide with Mars when the pericenter distance equals the Martian radius. Here, $a_{\rm R} \sim 9000$ km.}
\label{fig_life_shadow}
\end{figure}

Now, we define the \textit{time to collide} $\left(t_{\mathrm{col}}\right)$; the time it takes until a collision with Mars occurs. Using Eqs.~(\ref{eq_a_PR}) and (\ref{eq_a_sunshadow_lowest}), Fig.~\ref{fig_life_shadow} presents $t_{\mathrm{col}}$ for particles from $1a_{\rm R}$ (green) and $3a_{\rm R}$ (blue) in the case with (circle) and without (square) the planetary shadow. The black dashed line indicates the age of solar system (4 Gyr). The planetary shadow is responsible for longer lifetime for particles from $r_{\rm p}=10^{2} \mu \mathrm{m}$ to $r_{\rm p}=10 \mathrm{~m}$. Particles of $r_{\mathrm{p}} \lesssim 10 \mathrm{~cm}$ collide onto Mars within 4 Gyr in both cases with and without the planetary shadow.

Figure \ref{fig_life_shadow} indicates that the lifetime of particles is increased with the planetary shadow. For example, the particles of $10^{3} \mu m$ with the planetary shadow collide onto Mars in $2.46 \times 10^{7}$ years from $3a_{\rm R}$ (blue circle) and in $1.21 \times 10^{7}$ years from $1a_{\rm R}$ (green circle); when the planetary shadow is neglected, the collision happens in $1.38 \times 10^{7}$ years from $3a_{\rm R}$ (blue square) and $5.02 \times 10^{6}$ years from $1a_{\rm R}$ (green square). For these particles, the lifetime is increased by $81.97 \%$ starting at $3a_{\rm R}$ and by $139.88 \%$ starting at $1a_{\rm R}$. For other tested sizes, $t_{\rm col}$ is increased by 134.42\% ($10^{2} \mu m$), 69.82\% ($10^{4} \mu m$), 67.50\% ($10^{5} \mu \mathrm{m}$), 118.78\% ($1 \mathrm{m}$), and 95.42\% ($10 \mathrm{m}$) for particles starting at $3a_{\rm R}$ and by 113.81\% ($10^{2} \mu \mathrm{m}$), 94.98\% ($10^{4} \mu \mathrm{m}$), 65.97\% ($10^{5} \mu \mathrm{m}$), 66.63\% ($1 \mathrm{m}$), and 129.09\% ($10 \mathrm{m}$) starting at $1a_{\rm R}$.

Figure \ref{fig_life_shadow} (also Fig.~\ref{fig_shadow}) is based on analytical approximations. From numerical simulations, for example, \cite{Mak05} showed that $30 \mu m$ particles hit Mars within $4.70 \times 10^{5}$ years after it is released at $\sim 3a_{\rm R}$. Our evaluated lifetime for this particle is $10.36 \%$ larger than their numerical results. Figure \ref{fig_life_shadow} also shows that particles of $r_{\mathrm{p}} \lesssim 100 \mu \mathrm{m}$ from both at $1a_{\rm R}$ and $3a_{\rm R}$ collide with Mars within a timescale of $10^{5}$ years, which agrees with the numerical results of \cite{Liu20}. Thus, our analytical estimations provide an upper limit compared to the numerical simulations for various particle sizes, and can be used to conservatively estimate the dynamical lifetime of particles.

In short conclusion, under Martian $J_{2}$, the SRP, the PR force and the planetary shadow, particles up to $10^{5} \mu m$ ($=10 \mathrm{~cm}$) would eventually fall onto Mars with less than the age of the solar system. The main dynamical mechanism responsible for the decay in the pericenter distance depends on the particle size: small particles ($1 \mu m$ and $10 \mu m)$ are mainly regulated by Martian $J_{2}$ and the SRP, resulting in a rapid drop (within $\sim 1$ year), while large particles ($r_{\mathrm{p}}>10^{2} \mu m$) suffer from the cumulative effects of the PR force and the planetary shadow, spiraled onto Mars with a timescale of $\sim 10^{9}$ years. Although the particles up to $10^{5} \mu \mathrm{m}$ (=$10$ cm) collide with Mars before $4 \mathrm{Gyr}$ in both cases with and without the planetary shadow, the planetary shadow effect extends $t_{\rm col}$.

\subsection{Dependence on Martian obliquity}
\label{subsec:depend obliquity}

In Sects.~\ref{subsec:simu for J2&SRP&PR} and ~\ref{subsec:simu with sun} and Appendix \ref{sec:PR}, we derived an approximation of $\left<\frac{\mathrm{d} a}{\mathrm{d} t}\right>_{F_{\mathrm{PR}}}^{\epsilon \sim 0}$ (Eq.~(\ref{eq_PR_final}) and Eq.~(\ref{eq_a_PR})) with the assumption of small obliquity (regardless of $\Omega$), where we assumed $\cos ^{2} \epsilon \approx 1$ and $\sin ^{2} \epsilon \approx$ 0. The Martian obliquity $\epsilon$ is currently $\sim 25^{\circ}$, but is believed to be changed dramatically over billions of years \citep{War73,Tou93,Las93,Las04}. Our analytical solutions of the decay rate of the semi-major axis (also $p_{\max}$) explicitly contain the Martian obliquity $\epsilon$ (Fig.~\ref{fig_SunMars_conf}). In this subsection, we discuss the dependence on the Martian obliquity in the cases with and without the planetary shadow.

Without the planetary shadow, the decay of the semi-major axis (also $p_{\max}$) due to the PR force is mainly determined by $\boldsymbol{F}_{\mathrm{PR}, 4}$ and $\boldsymbol{F}_{\mathrm{PR}, 5}$ (see Appendix \ref{sec:PR}). Only the term $\left<\frac{\mathrm{d} a}{\mathrm{d} t}\right>_{F_{\mathrm{PR}, 5}}$ contains $\epsilon$ as follows (Eq.~(\ref{eq_PR5_final})),
\begin{equation}
\begin{aligned}
    \left<\frac{\mathrm{d} a}{\mathrm{d} t}\right>_{F_{\mathrm{PR}, 5}}  = -\frac{2}{n} \frac{B|\boldsymbol{V}|}{c|\boldsymbol{d}|^{2}} & \frac{1}{4}\left(\cos ^{2} \Omega+\cos ^{2} i \sin ^{2} \Omega + \cos ^{2} \epsilon \sin ^{2} \Omega \right. \\
     & \left. + \cos ^{2} i \cos ^{2} \Omega \cos ^{2} \epsilon+\sin ^{2} i \sin ^{2} \epsilon+\cos \epsilon \sin \epsilon \cos 2 i \cos \Omega \right).
\label{eq_obliquity_PR5}
\end{aligned}
\end{equation}
Then, Eq.~(\ref{eq_obliquity_PR5}) can be further simplified as
\begin{equation}
    \left<\frac{\mathrm{d} a}{\mathrm{d} t}\right>_{F_{\mathrm{PR}, 5}}=-\frac{2}{n} \frac{B|\boldsymbol{V}|}{c|\boldsymbol{d}|^{2}} \cdot \frac{1}{4}\left(\left(1+\cos ^{2} \Omega\right) \sin ^{2} i \sin ^{2} \epsilon+\cos ^{2} i+\cos ^{2} \epsilon+\frac{1}{2} \sin 2 \epsilon \cos ^{2} i \cos \Omega\right).
\label{eq_obliquity_PR5_final}
\end{equation}
In this paper, we set that the initial inclination of the particle is zero and assumed that the value of inclination is small to approximate the secular perturbations of the $\mathrm{PR}$ force and the effect of the planetary shadow. Thus, $\sin ^{2} i$ is a small value (much smaller than $\cos ^{2} i$). Then, the value of $\left<\frac{\mathrm{d} a}{\mathrm{d} t}\right>_{F_{\mathrm{PR}, 5}}$ (Eq.~(\ref{eq_obliquity_PR5_final})) mainly depends on the term $\left(\cos ^{2} i+\cos ^{2} \epsilon+\frac{1}{2} \sin 2 \epsilon \cos ^{2} i \cos \Omega\right)$. The minimum of $\left|\left<\frac{\mathrm{d} a}{\mathrm{d} t}\right>_{F_{\mathrm{PR}, 5}}\right|$ is achieved when the nodal line aligns with the $x$-axis of Mars-centered inertial frame, i.e., $\Omega=180^{\circ}$ ($\cos \Omega=-1$).

Thus, for a particle moving around Mars with arbitrary orientation of the nodal line, the value of $\left<\frac{\mathrm{d} a}{\mathrm{d} t}\right>{}_{F_{\mathrm{PR}, 5}}$ is constrained by the minimum of $\left|\left<\frac{\mathrm{d} a}{\mathrm{d} t}\right>_{F_{\mathrm{PR}, 5}} \right|$ with $\Omega=180^{\circ}$ as follows,
\begin{equation}
    \left<\frac{\mathrm{d} a}{\mathrm{d} t}\right>_{F_{\mathrm{PR}, 5}} \leq-\frac{2}{n} \frac{B|\boldsymbol{V}|}{c |\boldsymbol{d}|^{2}} \cdot \frac{1}{4}\left(\cos ^{2} i\left(1-\frac{1}{2} \sin (2 \epsilon)\right)+\cos ^{2} \epsilon\right)<0
\label{eq_obliquity_PR5_maximum}
\end{equation}
Summarizing $\left<\frac{\mathrm{d} a}{\mathrm{d} t}\right>_{F_{\mathrm{PR},5}}$ (Eq.~(\ref{eq_obliquity_PR5_maximum})) and $\left<\frac{\mathrm{d} a}{\mathrm{d} t}\right>_{F_{\mathrm{PR},4}}$ (Eq.~(\ref{eq_PR4_integ})) yields the lowest decay rate outside the planetary shadow, $\left<\frac{\mathrm{d} a}{\mathrm{d} t}\right>_{F_{\mathrm{PR},\rm lowest}}^{\rm no-shadow}$, as
\begin{equation}
     \left<\frac{\mathrm{d} a}{\mathrm{d} t}\right>_{F_{\mathrm{PR}}} = \left<\frac{\mathrm{d} a}{\mathrm{d} t}\right>_{F_{\mathrm{PR},4}} + \left<\frac{\mathrm{d} a}{\mathrm{d} t}\right>_{F_{\mathrm{PR},5}} \leq \left<\frac{\mathrm{d} a}{\mathrm{d} t}\right>_{F_{\mathrm{PR},\rm lowest}}^{\rm no-shadow} ,
\label{eq_obliquity_PR_inequality}
\end{equation}
where 
\begin{equation}
    \left<\frac{\mathrm{d} a}{\mathrm{d} t}\right>_{F_{\mathrm{PR},\rm lowest}}^{\rm no-shadow} = -\frac{2}{n} \frac{B|\boldsymbol{V}|}{c|\boldsymbol{d}|^{2}}\left[1+\frac{1}{4}\left(\cos ^{2} i\left(1-\frac{1}{2} \sin (2 \epsilon)\right)+\cos ^{2} \epsilon \right) \right] < 0 .
\label{eq_obliquit_PR_lowest}
\end{equation}
Here, we approximated the decay rate of the semi-major axis assuming that particle's nodal line aligns with the $-x$-axis of Mars-centered inertial frame (Eq.~(\ref{eq_obliquit_PR_lowest})). We note that particles with arbitrary orientation of the nodal line (or $\Omega$) would have smaller semi-major axis than the one estimated by this treatment.

Below, we consider a particle around Mars with Martian $J_{2}$, the solar radiation forces, and the planetary shadow. When the semi-major axis of a particle is initially large enough with a non-zero $\epsilon$, its initial orbit could be entirely out of the planetary shadow. The semi-major axis, then, may be gradually decreased by the PR force. When the semi-major axis decreases to the critical orbital distance, i.e., $a_{\text {intersect}}=R_{\mathrm{M}} / \sin \epsilon$, the particle has an orbit that geometrically intersects the shadow cylinder (see Fig.~\ref{fig_SunMars_conf}(a)). The semi-major axis further decreases by the PR force with the planetary shadow effect.

Thus, for a nonzero $\epsilon$, when $a>a_{\text {intersect}}$ (or $p_{\max}>a_{\text {intersect}}$), the evolution of $p_{\max}$ is described by Eq.~(\ref{eq_obliquit_PR_lowest}) (i.e., no planetary shadow), while, for $a<a_{\text {intersect}}$ (or $p_{\max}<a_{\text {intersect}}$), the planetary shadow effect is considered based on Eq.~(\ref{eq_a_sunshadow_lowest}). The values of $a_{\text {intersect}}$ with $\epsilon=25^{\circ}, 45^{\circ}$, and $75^{\circ}$ are $0.88 a_{\mathrm{R}}, 0.53 a_{\mathrm{R}}$, and $0.38 a_{\mathrm{R}}$, respectively. A unique situation occurs when $\epsilon=90^{\circ}$ because the particle's entire orbit is outside the shadow cylinder for any given value of the semi-major axis, meaning that the inclusion of the planetary shadow does not influence its motion.

As explained above, in the case with the planetary shadow, we use Eq.~(\ref{eq_a_sunshadow_lowest}). The lowest decay rate from Eq.~(\ref{eq_a_sunshadow_lowest}) is realized when $\cos \epsilon=1$ (i.e., the value of $\epsilon$ is small). For an arbitrary non-zero value of $\epsilon$, the decay rate of the semi-major axis is higher than $\cos \epsilon=1$ case. Thus, their collision time onto Mars is shorter than $\cos \epsilon=1$ case.

\begin{figure}
    \centering
    \includegraphics[width=0.7\textwidth]{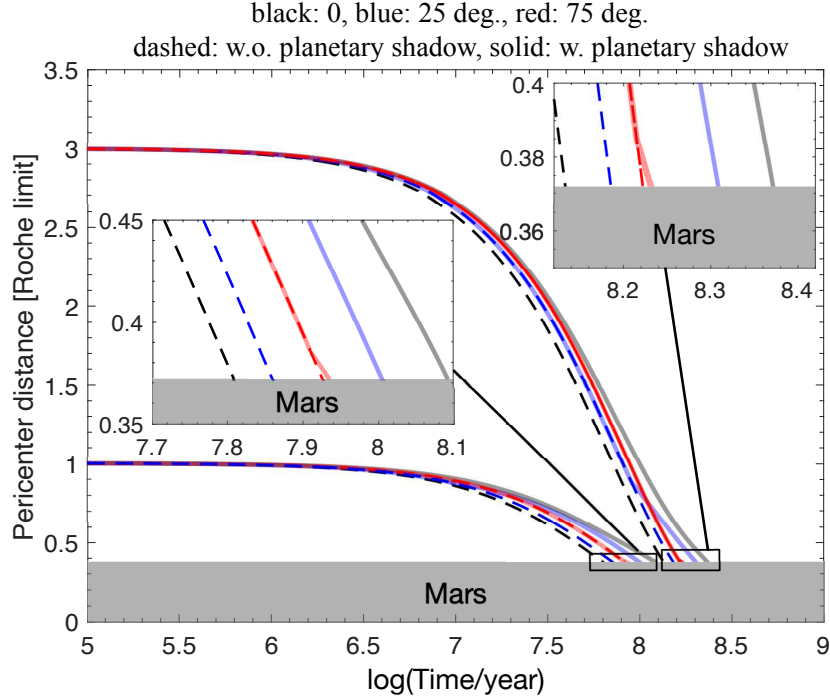}
    \caption{Analytical estimations of the pericenter distance for different values of obliquity $\epsilon$. The solid lines indicate $p_{\rm max}$ with the planetary shadow. The dashed lines indicate those without the planetary shadow. The blue solid lines overlap with the blue dashed lines when $p_{\rm max}$ is larger than $0.88 a_{\rm R}$ ($a_{\rm intersect}$ of $\epsilon=25^{\circ}$); The red solid lines overlap with the red dashed lines when $p_{\rm max}$ is larger than $0.38 a_{\rm R}$ ($a_{\rm intersect}$ of $\epsilon=75^{\circ}$); the black solid lines do not overlap with the black dashed lines for $\epsilon=0^{\circ}$. All tested particles collide onto the Martian surface within $10^9$ years. Here, $a_{\rm R} \sim 9000$ km.}
\label{fig_obliquity}
\end{figure}

Figure \ref{fig_obliquity} shows the analytical estimations of the pericenter distance (i.e., $p_{\max}$) for a particle of $1 \mathrm{~cm}$ as $\epsilon=0^{\circ}$ (black), $25^{\circ}$ (blue), and $75^{\circ}$ (red) starting from $1 a_{\mathrm{R}}$ and $3 a_{\mathrm{R}}$ in both cases with (solid) and without (dashed) planetary shadow. As presented by the zoom-in panels of dashed lines, in the case without the planetary shadow, the dependency of the obliquity on the pericenter distance evolution is relatively weak. 

When we include the planetary shadow, the red and blue solid lines overlap with those dashed lines before $a$ decreases to the corresponding critical distance, i.e., $a=0.38 a_{\mathrm{R}}$ and $0.88 a_{\mathrm{R}}$, respectively. As illustrated by the solid lines, with the obliquity increasing, the decay of the semi-major axis becomes faster because of a less shadowed portion during the orbit. However, the lifetime of $1 \mathrm{~cm}$-sized particle is not significantly changed after including a non-zero obliquity with the planetary shadow and the particle collides onto Mars within $10^{9}$ years.

\begin{figure}
    \centering
    \includegraphics[width=\textwidth]{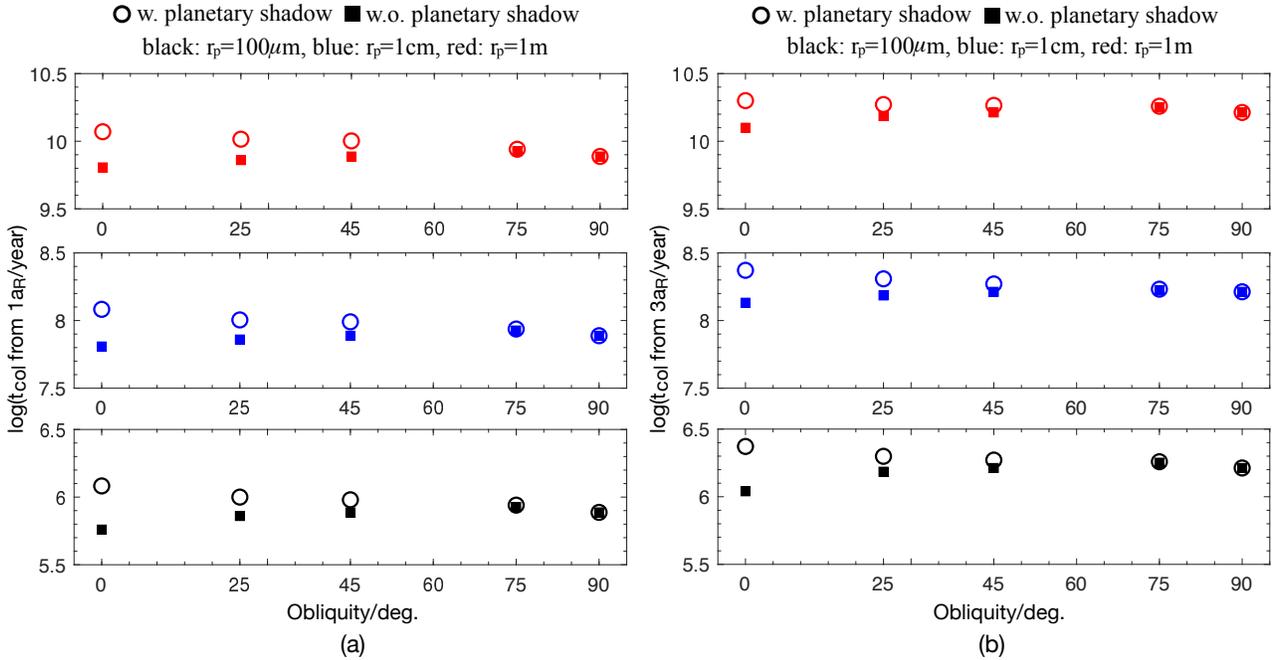}
    \caption{Time to collide for particles starting from $1 a_{\rm R}$ (panel (a)) and $3 a_{\rm R}$ (panel (b)) for different values of obliquity $\epsilon$ in the case with (open circles) and without (filled squares) the planetary shadow. The particle sizes vary between $100 \mu {\rm m}$ (black), $1 {\rm cm}$ (blue), and $1 {\rm m}$ (red). For all tested obliquity in both cases with and without the planetary shadow, particles up to $\sim 1 {\rm cm}$ collide onto Mars within $10^9$ years, while meter-sized particles do not collide within $4$ Gyr. Here, $a_{\rm R} \sim 9000$ km.}
\label{fig_life_obliquity}
\end{figure}

Figure \ref{fig_life_obliquity} illustrates the lifetime of particles of $100 \mu \mathrm{m}$ (black), $1 \mathrm{~cm}$ (blue), and $1 \mathrm{~m}$ (red) starting from 1 $a_{\mathrm{R}}$ (a) and $3 a_{\mathrm{R}}$ (b) for different values of obliquity $\epsilon$. Both cases with and without the planetary shadow are discussed. Without the planetary shadow, a non-zero value of $\epsilon$ causes little effect in the lifetime. A slight increase in the lifetime is caused as the value of $\epsilon$ increases from $0^{\circ}$ to $90^{\circ}$. The time to collide for all non-zero $\epsilon$ is within $34 \%$ larger than the result of $\epsilon=0^{\circ}$.

When we include the planetary shadow, the lifetime of particles decreases as $\epsilon$ increases from $0^{\circ}$ to $90^{\circ}$ (with $37 \%$ decrement in lifetime at most). The black solid lines never overlap with the black dashed lines for $\epsilon=0^{\circ}$. This is because a larger fraction of particle's orbit is outside the shadow cylinder as obliquity increases, leading to a faster decrease in the semi-major axis due to the PR force. We note that the orbits of particles are entirely outside the planetary shadow when their orbital distances are large enough. Including this effect, for all the values of $\epsilon$, particles of $r_{\mathrm{p}} \lesssim 10 \mathrm{~cm}$ starting at both Phobos and Deimos orbits collide with Mars within $\sim 4 \mathrm{Gyr}$.

In short conclusion, without the planetary shadow, the Martian obliquity slightly changes the lifetime of a particle by about $34 \%$ at most. With the planetary shadow, a particle collides onto Mars faster for a non-zero obliquity (with $37 \%$ decrement in lifetime at most). However, in both cases with the inclusion of the Martian obliquity (from $\epsilon=0^{\circ}$ to $90^{\circ}$), the conclusion we obtained in Sec.~\ref{subsec:compar wo sun} still holds that particles up to $10^{5} \mu m$ (=10 cm) eventually fall onto Mars with less than the age of the solar system.

\subsection{Comparison to Rubincam (2013)} 
\label{subsec:Rubincam}

\cite{Rub13} assumed zero-obliquity of planet $(\epsilon=0)$. Substituting $\epsilon=0$ to Eq.~(\ref{eq_PR3_final}) yields
\begin{equation}
    \left<\frac{\mathrm{d} a}{\mathrm{d} t}\right>_{F_{\mathrm{PR},3}}=-\frac{2}{n} \frac{B\left|\boldsymbol{V}_{\mathrm{S}}\right|\left|\boldsymbol{R}_{\mathrm{S}}\right||\boldsymbol{R}|}{c|\boldsymbol{d}|^{4}}\left(\frac{1}{2} \cos i\right) ,
\label{eq_Rubincam_PR3}
\end{equation}
which coincides with the Eq. (27) of \cite{Rub13}. Equation (\ref{eq_PR5_final}) can be further simplified as
\begin{equation}
    \left<\frac{\mathrm{d} a}{\mathrm{d} t}\right>_{F_{\mathrm{PR},5}}=-\frac{2}{n} \frac{B|\boldsymbol{V}|}{c|\boldsymbol{d}|^{2}} \frac{1+\cos ^{2} i}{4} ,
\label{eq_Rubincam_PR5}
\end{equation}
which agrees with Eq. (32) of \cite{Rub13}. Note that $\boldsymbol{F}_{\mathrm{PR},1}$, $\boldsymbol{F}_{\mathrm{PR},2}$ and $\boldsymbol{F}_{\mathrm{PR},6}$ are negligible \citep{Rub13}. So far, the mathematical deductions on the secular perturbation on $a$ due to the PR effect match those in \cite{Rub13}. 

\cite{Rub13} also considered the planetary shadow, suggesting that the shadowed SRP alone exerts a negative effect on the variation of $a$, referring to Eq. (21) of \cite{Rub13}. However, there is an error in the mathematical derivation at Eq. (20) of \cite{Rub13}. In Eq. (20) of \cite{Rub13}, the tangential component of the SRP force of the term contains an error as explained below.

In Eq. (20) of \cite{Rub13}, $\hat{\boldsymbol{y}}$ indicates the orientation of the $y$-axis of a Mars-rotational frame, thus it indicates the direction of the sunlight towards Mars (i.e., from the Sun to Mars). The coordinates of $\hat{\boldsymbol{y}}$ which \cite{Rub13} used were expressed in the Mars-rotational frame, while those for $\hat{\boldsymbol{t}}$ were shown in a Mars-centered inertial frame. Such inconsistency in the coordinates led to an error. This error is nonnegligible because the basic frame used in the related mathematical deduction contains an error. For the same reason, the conclusion in \cite{Rub13} -- saying that the net orbital evolution in $a$ due to the sum of $\boldsymbol{F}_{\mathrm{PR},1}$ and $\boldsymbol{F}_{\rm SRP}$ with the planetary shadow is zero -- is incorrect.

In Appendix \ref{sec:SRP and J2} and \cite{Hau13}, the numerical results showed that the shadowed SRP alone exerts a periodic effect on the variation of $a$, resulting in a long-term periodic variation, which is different from the conclusion of Eq. (21) of \cite{Rub13}, i.e., the negative effect of the shadowed SRP in the semi-major axis.

\section{Discussion}
\label{sec:diss}
\subsection{Collision with Phobos and Deimos}
\label{subsec:collision}

In this study, we did not include Phobos and Deimos in our numerical simulations. After the formation of Phobos and Deimos, however, continuous meteoroidal impacts occur on these moons and dust would be ejected from the surface of the moons \citep[e.g.,][]{Ram13,Kur19,Hyo19}. The produced dust particles around Mars initially have orbits that cross the moon (because it is the launch point) and some would re-accrete during the successive orbits. 

\cite{Liu20} indeed performed a direct numerical integration of particles (grain sizes ranging from $0.5$ to $100$ $\mu$m) that are ejected from Phobos and Deimos. They considered the re-accretion onto the Martian moons. They showed that small particles ($\lesssim 10$ $\mu$m) hit Mars just after it is released \cite[lifetime less than a year; see also][]{Ham96}. They also reported that larger particles stay in Phobos orbit for only $\sim 10-100$ years and Deimos orbit for $\sim 10^4$ years or more before they hit the source moon or Mars. Their shorter lifetime of larger particles (e.g. $r_{\rm p} \sim 100~\mu$m) compared to ours (e.g. Fig.~\ref{fig_life_shadow}) is still consistent. This is because larger particles take more time to change their orbits, during which these large particles efficiently accrete onto Phobos or Deimos as observed in \cite{Liu20}, while our study did not include Phobos and Deimos.

Our study focused on the dynamical evolution of particles by Martian gravity, Martian $J_2$, and non-gravitational forces. Depending on how dust particles are formed, collisions with Phobos and Deimos could be responsible for removing particles around Mars. Therefore, the results of our study would be the upper limit of the dynamical lifetime of particles that ignores such an additional dynamical process.

\subsection{On the origin of Phobos and Deimos in giant impact hypothesis}
\label{subsec:impact}

Various scenarios have been proposed regarding the giant impact origin of Martian moons -- i.e., various initial masses of the debris disk, followed by different disk evolutions -- are proposed \citep[e.g.,][]{Cra11,Ros16,Can18,Hes17,Bag21,Hyo17a,Hyo17b,Hyo18,Hyo22}. The debris disk evolves by the self-gravity and collision among constituent particles until the disk becomes thin enough. As a natural consequence of the giant impact hypothesis, after the formation of Phobos and Deimos, a thin remnant particulate disk is expected to be left behind.

Today's observation around Mars, however, did not report an obvious particle disk around Mars with optical depth $\tau>3 \times 10^{-5}$ \citep{Dux88}. \cite{Sho06} did not detect a particle around Mars down to the detection limit of $75\mathrm{~m}$. Therefore, if the giant impact hypothesis is correct, some mechanism must be responsible for removing the remnant disk particles.

Importantly, depending on the giant impact hypothesis, the timing of the formation of Phobos -- that is, the timing of the remnant disk formation -- is significantly different. For example, \cite{Hes17} considered an ancient giant impact ($>4.3$ Gyr ago) and proposed successive several of the disk-moon recycling evolution that continues until Phobos is recently formed (hundreds of millions of years ago, assuming disk particles have $r_{\rm p} \sim 10 \mathrm{~cm}$), indicating that this scenario requires the removal of the remaining disk particles ($r_{\rm p} \sim 10 \mathrm{~cm}$) on a timescale of hundreds of millions of years. On the other hand, \cite{Ros16,Hyo17a,Can18} proposed formation of Phobos and Deimos several Gyrs ago. 

Our study showed that particles of $r_{\rm p} \sim 10 \mathrm{~cm}$ can be removed by the cumulative effects of the PR drag to Mars over $\sim 10^{9}$ years; smaller particles require less time to be removed, and vice versa. Thus, it seems unlikely that the PR drag (and $J_{2}$ and the SPR forces) alone can remove the remaining particle disk in the context of \cite{Hes17} and another process is needed to remove the particles to validate their model. If Phobos formed billions years ago and a thin particle disk is left behind, the remaining particles smaller than $r_{\rm p} \sim 10 \mathrm{~cm}$ could be removed by the cumulative effect of the PR force over $\sim 10^{9}$ years.

Further observational constrains of today's putative remaining particles around Mars (or observational more solid evidence, for example, reporting no particle around Mars) together with our theoretical results may validate and/or refute some of the giant impact scenarios to better constrain the origin of Phobos and Deimos. Future planetary mission would also play an important role.

\section{Summary}
\label{sec:summary}

In this paper, using a direct numerical integration and analytical arguments, we studied the giga-year dynamical evolution of particles around Mars starting at Phobos and Deimos orbits under various perturbations: Martian $J_{2}$, the solar radiation pressure (SRP), and the Poynting-Robertson (PR) drag force. We also studied the effects of the planetary shadow.

The PR force exerts a cumulative effect and the semi-major axis of a particle decays over time. The decay rate strongly depends on the particle size. Previous studies did not investigate the effect of the planetary shadow for giga-year timescale. We showed that the planetary shadow changes the lifetime of large particles ($r_{\mathrm{p}} \gtrsim 10 \mu m$), in that, it extends the decay timescale of particle orbits. We derived an analytical upper bound of the pericenter distance $\left(p_{\max}\right)$, which agreed with the numerical results and was used to estimate the upper limit of the dynamical lifetime of inward drifting particles (due to the PR force) until the collision with Mars occurs. Our main findings are as follows:

\begin{itemize}
\setlength{\parskip}{0cm} 
\setlength{\itemsep}{0.3cm}

 \item[--] Particles smaller than $r_{\mathrm{p}} \sim 10 \mu m$ starting at Phobos and Deimos orbits are dominated by the SRP and collide with Mars within $\sim 1$ year as the eccentricities are pumped up (see also \cite{Kri96} and \cite{Sas99}). Thus, small particles $\left(r_{\mathrm{p}} \lesssim 10 \mu m\right)$ at Phobos and Deimos orbits are expected to be rapidly removed, which was also reported by \cite{Mak05} and \cite{Liu20}.

 \item[--] Large particles $\left(r_{\mathrm{p}}>10 \mu m\right)$ have longer lifetime: those at Phobos and Deimos orbits can survive more than $10^{4}$ years until they spiral onto Mars. Here, the cumulative effect of the PR force is responsible for the decay in the semi-major axis and thereby in the pericenter distance.

 \item[--] For particles at Deimos orbit with sizes from $r_{\mathrm{p}}=30 \mu m$ to $r_{\mathrm{p}}=1000 \mu m$, changing initial eccentricity between $0.1$ and $0.7$ does not lead to a rapid collision with Mars (Sec.~\ref{subsec:depend size}). The same conclusion was obtained for particles of $r_{\mathrm{p}}=1000 \mu m$ at Phobos orbit.

 \item[--] When the planetary shadow is considered, the lifetime of particles is extended (Fig.~\ref{fig_life_shadow}). For example, the lifetime of particles of $r_{\mathrm{p}}=10^{2} \mu m$ at Phobos and Deimos orbits becomes $113.81 \%$ and $134.42 \%$ longer than the cases without the planetary shadow. Even including the planetary shadow, particles up to $r_{\mathrm{p}} \sim 10 \mathrm{~cm}$ at $\lesssim 8$ Martian radius eventually spiral onto the Martian surface within $\sim 10^{9}$ years.
 
 \item[--] In the case with and without the planetary shadow, the Martian obliquity slightly changes the lifetime of particles (Sec.~\ref{subsec:depend obliquity}). For example, without the planetary shadow, the lifetime of particles increases by about $34 \%$ at most as the Martian obliquity increases from 0 to $90^{\circ}$. Regardless of the value of the obliquity, particles up to $r_{\mathrm{p}} \sim 10 \mathrm{~cm}$ at $\lesssim 8$ Martian radius eventually fall onto Mars within $\sim 10^{9}$ years.

\end{itemize}

Although some uncertainties remain, e.g., the radiation pressure efficiency $Q$ (where we assumed $Q=1$) and particle density (where we assumed $3 \, \mathrm{g} \, \mathrm{cm}^{-3}$), our study would provide key constrains regarding the origin and evolution of Martian moons, Phobos and Deimos.

Martian Moons eXploration (MMX) mission, developed by the Japan Aerospace eXploration Agency (JAXA), is expected to be launched in 2024 with the aims of elucidating the origin of Martian moons \citep{Fuj19}, collecting geochemical information about the evolution of Martian surface environment \citep{Hyo19}, and searching for traces of Martian life \citep{Hyo21}. A Circum-Martian Dust Monitor is scheduled to be onboard MMX \citep{Kob18}. Together with data collected by the MMX mission, we hope our results will help shed light on the origin and evolution of Martian moon systems.

\appendix

\section{Analytic Estimation of Particle Evolution}
\label{sec:analytic}

To evaluate the perturbation of external forces upon a particle in a Keplerian orbit, we use the Planetary Perturbation Equation (see e.g., \cite{Mur99}). The Lagrange's equations with the disturbing potential $\bar{R}$ on the semi-major axis $a$ and the eccentricity $e$ are
\begin{equation}
\begin{aligned}
    &\left<\frac{\mathrm{d} a}{\mathrm{d} t}\right>=\frac{2}{n a} \frac{\partial \bar{R}}{\partial f}
\end{aligned}
\label{eq_Lagrange_a}
\end{equation}
and
\begin{equation}
\begin{aligned}
    &\left<\frac{\mathrm{d} e}{\mathrm{d} t}\right>=\frac{1-e^{2}}{n a^{2} e} \frac{\partial \bar{R}}{\partial f}-\frac{\sqrt{1-e^{2}}}{n a^{2} e} \frac{\partial \bar{R}}{\partial \omega},
\end{aligned}
\label{eq_Lagrange_e}
\end{equation}
where $f$ is the true anomaly, $n$ is the mean angular velocity, and $\omega$ is the argument of pericenter.

We also use Gauss's equations for the variation of the orbital elements, which are expressed in terms of the perturbing acceleration. When $F_{\mathrm{U}}$, $F_{\mathrm{N}}$, and $F_{\mathrm{B}}$ denote the tangential (or along-track), the normal, and the binormal components of an external force $\boldsymbol{F}$ exerted upon a particle, respectively, the equations for the variation of Kepler orbital elements ($a$, $e$, $i$, $\omega$, $\Omega$, $f$) can be obtained. Taking $a$ and $e$ as an example, their variations are \citep{Bla61}

\begin{equation}
\begin{gathered}
\frac{\mathrm{d} a}{\mathrm{d} t}=\frac{2}{n} F_{\mathrm{U}}
\end{gathered}
\label{eq_a_Force}
\end{equation}
and
\begin{equation}
\begin{gathered}
\frac{\mathrm{d} e}{\mathrm{d} t}=\frac{1}{v}\left(2(e+\cos f) \cdot F_{\mathrm{U}}+\frac{1}{a} r \sin f \cdot F_{\mathrm{N}}\right) ,
\end{gathered}
\label{eq_e_Force}
\end{equation}
where $r$ is the distance to the center of the planet and $v$ is the velocity. Implementing averaging method to Eq.~(\ref{eq_a_Force}) and Eq.~(\ref{eq_e_Force}), the secular evolution of $a$ and $e$ can be estimated mathematically as
\begin{equation}
\begin{gathered}
    \left<\frac{\mathrm{d} a}{\mathrm{d} t}\right>=\frac{1}{2 \pi} \frac{2}{n} \int_{0}^{2 \pi} F_{\mathrm{U}}(f) \mathrm{d} f \\
\end{gathered}
\label{eq_secu_a}
\end{equation}
and
\begin{equation}
\begin{gathered}
    \left<\frac{\mathrm{d} e}{\mathrm{d} t}\right>=\frac{1}{2 \pi} \int_{0}^{2 \pi} \frac{1}{v}\left(2(e+\cos f) \cdot F_{\mathrm{U}}+\frac{1}{a} r \sin f \cdot F_{\mathrm{N}}\right) \mathrm{d} f .
\end{gathered}
\label{eq_secu_e}
\end{equation}
In this study, using the averaging method, we explain the resultant contributions arising from Martian $J_{2}$, the SRP, the $\mathrm{PR}$, and the planetary shadow effects, respectively.

\section{Poynting-Robertson force}
\label{sec:PR}

Following \cite{Rub13} and \cite{Bur79}, the PR force, $\boldsymbol{F}_{\mathrm{PR}}$, is decomposed to 6 terms concerning relative motions with respect to Mars and the motion of Mars around the Sun as follows:
\begin{equation}
    \boldsymbol{F}_{\rm PR}=\frac{B}{|\boldsymbol{d}|^{2}}\left[-\frac{\boldsymbol{V}_{\mathrm{S}}}{c}-\frac{\boldsymbol{V}_{\mathrm{S}} \cdot \boldsymbol{R}_{\mathrm{S}}}{c|\boldsymbol{d}|^{2}} \boldsymbol{d}-\frac{\boldsymbol{V}_{\mathrm{S}} \cdot \boldsymbol{R}}{c|\boldsymbol{d}|^{2}} \boldsymbol{d}-\frac{\boldsymbol{V}}{c}-\frac{\boldsymbol{V} \cdot \boldsymbol{R}_{\mathrm{S}}}{c|\boldsymbol{d}|^{2}} \boldsymbol{d}-\frac{\boldsymbol{V} \cdot \boldsymbol{R}}{c|\boldsymbol{d}|^{2}} \boldsymbol{d}\right]=\sum_{i=1}^{6} \boldsymbol{F}_{{\rm PR}, i},
\label{eq_PR_each}
\end{equation}
where $\boldsymbol{F}_{{\rm PR}, i}(i=1,2, \ldots, 6)$ denotes each term in $\boldsymbol{F}_{\rm PR}$ from left to right. $\boldsymbol{d}$ is the position vector of the particle with respect to the Sun. $\boldsymbol{R}$ denotes the position vector of the particle with respect to the Mars and $\boldsymbol{R}_{\mathrm{S}}$ is the position vector of Mars with respect to the Sun. $\boldsymbol{V}$ denotes the velocity vector of the particle with respect to Mars and $\boldsymbol{V}_{\mathbf{S}}$ is the velocity vector of Mars with respect to the Sun. Thus, $\boldsymbol{d}=\boldsymbol{R}+\boldsymbol{R}_{\mathrm{S}}$. In the following texts, the vectors, these vectors are cast into the Mars-centered inertial $(\mathrm{O}-x-y-z)$ frame where their coordinates are presented. 

To deduce the secular perturbation of PR effect, we set the following assumption: $(i, \omega, \Omega)$ remain constant in one period and $e \simeq 0$ in one period (for its initial value is 0). Under such assumption, several important vectors in the $(\mathrm{O} - \left.x-y-z\right)$ frame can be mathematically expressed as follows. The expressions of $\left(\boldsymbol{R}_{\mathrm{S}}, \boldsymbol{V}_{\mathrm{S}}\right)$ are
\begin{equation}
    \boldsymbol{R}_{\mathrm{S}}=\left|\boldsymbol{R}_{\mathrm{S}}\right| \cdot\left(\cos n_{\mathrm{M}} t \cdot \boldsymbol{i}_{x}+\cos \epsilon \cdot \sin n_{\mathrm{M}} t \cdot \boldsymbol{i}_{y}+\sin \epsilon \cdot \sin n_{\mathrm{M}} t \cdot \boldsymbol{i}_{z}\right)
\label{eq_Rs}
\end{equation}
and
\begin{equation}
    \boldsymbol{V}_{\mathrm{S}}=\left|\boldsymbol{V}_{\mathrm{S}}\right| \cdot\left(-\sin n_{\mathrm{M}} t \cdot \boldsymbol{i}_{x}+\cos \epsilon \cdot \cos n_{\mathrm{M}} t \cdot \boldsymbol{i}_{y}+\sin \epsilon \cdot \cos n_{\mathrm{M}} t \cdot \boldsymbol{i}_{z}\right) ,
\label{eq_Vs}
\end{equation}
where $n_{\mathrm{S}}$ denotes the mean angular velocity of Mars around the Sun and $n_{\mathrm{M}}=\frac{2 \pi}{T_{\mathrm{M}}}$. ($\boldsymbol{i}_{x}$, $\boldsymbol{i}_{y}$, $\boldsymbol{i}_{z}$) are the unit vectors of three axes in the (O$-x-y-z$) frame. Given that $e \simeq 0$, the tangential unit vector $\hat{\boldsymbol{t}}$ and the radial unit vector $\hat{\boldsymbol{r}}$ satisfy:
\begin{equation}
    \hat{\boldsymbol{t}}=(-\sin u \cos \Omega-\cos u \sin \Omega \cos i) \cdot \boldsymbol{i}_{x}+(-\sin u \sin \Omega+\cos u \cos \Omega \cos i) \cdot \boldsymbol{i}_{y}+(\cos u \sin i) \cdot \boldsymbol{i}_{z}
\label{eq_vec_t}
\end{equation}
and
\begin{equation}
    \hat{\boldsymbol{r}}=(\cos u \cos \Omega-\sin u \sin \Omega \cos i) \cdot \boldsymbol{i}_{x}+(\cos u \sin \Omega+\sin u \cos \Omega \cos i) \cdot \boldsymbol{i}_{y}+(\sin u \sin i) \cdot \boldsymbol{i}_{z},
\label{eq_vec_r}
\end{equation}
where $u=\omega+f$. The expressions of $(\boldsymbol{R}, \boldsymbol{V})$ cast into $(\mathrm{O}-x-y-z)$ frame are $\boldsymbol{R}=|\boldsymbol{R}| \cdot \hat{\boldsymbol{r}}$ and $\boldsymbol{V}=|\boldsymbol{V}| \cdot \hat{\boldsymbol{t}}$. Now, according to Eq.~(\ref{eq_secu_a}), the secular perturbation on $a$ is caused by the tangential component of each $\boldsymbol{F}_{{\rm PR}, i}$ and can be calculated as
\begin{equation}
    \left<\frac{\mathrm{d} a}{\mathrm{d} t}\right>_{F_{\mathrm{PR}, i}}=\frac{1}{2 \pi} \frac{2}{n} \int_{0}^{2 \pi} \boldsymbol{F}_{\mathrm{PR}, i}(f) \cdot \hat{\boldsymbol{t}} \mathrm{d} f .
\label{eq_PR_integ}
\end{equation}
Here, $\left<\frac{\mathrm{d} a}{\mathrm{d} t}\right>_{F_{\mathrm{PR}, 1}}$ is given as
\begin{equation}
    \left<\frac{\mathrm{d} a}{\mathrm{d} t}\right>_{F_{\mathrm{PR}, 1}}=-\frac{B}{c|\boldsymbol{d}|^{2}} \frac{1}{2 \pi} \frac{2}{n} \int_{0}^{2 \pi} \boldsymbol{V}_{\mathrm{S}} \cdot \hat{\boldsymbol{t}} \mathrm{d} f.
\label{eq_PR1_integ}
\end{equation}
Each term in the integral of Eq.~(\ref{eq_PR1_integ}) contains either $\sin$ or $\cos$, so its averaged value is zero. $\boldsymbol{F}_{{\rm PR}, 2}$ is given as
\begin{equation}
  \left<\frac{\mathrm{d} a}{\mathrm{d} t}\right>_{F_{\mathrm{PR}, 2}}=-\frac{B}{c|\boldsymbol{d}|^{4}} \frac{1}{2 \pi} \frac{2}{n} \int_{0}^{2 \pi} \boldsymbol{V}_{\mathrm{S}} \cdot \boldsymbol{R}_{\mathrm{S}} \cdot \boldsymbol{d} \cdot \hat{\boldsymbol{t}} \mathrm{d} f=0.
\label{eq_PR2_integ} 
\end{equation}
This is because Mars is assumed to orbit the Sun in a circular motion and $\boldsymbol{V}_{\mathrm{S}} \cdot \boldsymbol{R}_{\mathrm{S}}=0$.
$\boldsymbol{F}_{\mathrm{PR}, 3}$ is given as
\begin{equation}
    \left<\frac{\mathrm{d} a}{\mathrm{d} t}\right>_{F_{\mathrm{PR}, 3}}=-\frac{B}{c|\boldsymbol{d}|^{4}} \frac{1}{2 \pi} \frac{2}{n} \int_{0}^{2 \pi} \boldsymbol{V}_{\mathrm{S}} \cdot \boldsymbol{R} \cdot \boldsymbol{d} \cdot \hat{\boldsymbol{t}} \mathrm{d} f.
\label{eq_PR3_integ}
\end{equation}
To completely eliminate all periodic terms, double averaging upon the true anomaly and the argument of pericenter is implemented so that terms of $\cos n_{\mathrm{M}} t$ and $\sin n_{\mathrm{M}} t$ are eliminated. Thus,
\begin{equation}
    \left<\frac{\mathrm{d} a}{\mathrm{d} t}\right>_{F_{\mathrm{PR}, 3}}=\frac{2}{n} \frac{B\left|\boldsymbol{V}_{\mathrm{S}}\right|\left|\boldsymbol{R}_{\mathrm{S}}\right||\boldsymbol{R}|}{c|\boldsymbol{d}|^{4}}\left(\frac{1}{2} \cos i \cos \epsilon-\frac{1}{2} \sin i \sin \epsilon \cos \Omega\right).
\label{eq_PR3_final}
\end{equation}
$\boldsymbol{F}_{{\rm PR}, 4}$ is given as
\begin{equation}
    \left<\frac{\mathrm{d} a}{\mathrm{d} t}\right>_{F_{\mathrm{PR}, 4}}=-\frac{B}{c|\boldsymbol{d}|^{2}} \frac{1}{2 \pi} \frac{2}{n} \int_{0}^{2 \pi} \boldsymbol{V} \cdot \hat{\boldsymbol{t}} \mathrm{d} f=-\frac{2}{n} \frac{B|\boldsymbol{V}|}{c|\boldsymbol{d}|^{2}} .
\label{eq_PR4_integ}
\end{equation}
Equation (\ref{eq_PR4_integ}) contains a drag term because the force ($\boldsymbol{F}_{{\rm PR}, 4}$) acts against the velocity vector $(\boldsymbol{V})$. $\boldsymbol{F}_{{\rm PR}, 5}$ is given as
\begin{equation}
    \left<\frac{\mathrm{d} a}{\mathrm{d} t}\right>_{F_{\mathrm{PR}, 5}}=-\frac{B}{c|d|^{4}} \frac{1}{2 \pi} \frac{2}{n} \int_{0}^{2 \pi} \boldsymbol{V} \cdot \boldsymbol{R}_{\mathrm{S}} \cdot \boldsymbol{d} \cdot \hat{\boldsymbol{t}} \mathrm{d} f ,
\label{eq_PR5_integ}
\end{equation}
where
\begin{equation}
    \int_{0}^{2 \pi} \boldsymbol{V} \cdot \boldsymbol{R} \cdot \boldsymbol{d} \cdot \hat{\boldsymbol{t}} \mathrm{d} f=|\boldsymbol{V}| \int_{0}^{2 \pi}\left(\hat{\boldsymbol{t}} \cdot \boldsymbol{R_{\mathrm{S}}}\right) \cdot(\boldsymbol{d} \cdot \hat{\boldsymbol{t}}) \mathrm{d} f .
\label{eq_VRd}
\end{equation}
Assuming that $\boldsymbol{d} \simeq \boldsymbol{R}_{\rm{S}}$,
\begin{equation}
\begin{aligned}
    \left<\frac{\mathrm{d} a}{\mathrm{d} t}\right>_{F_{\mathrm{PR}, 5}}=-\frac{2}{n} \frac{B|\boldsymbol{V}|}{c|\boldsymbol{d}|^{2}} & \frac{1}{4} \left( \cos^{2} \Omega+\cos^{2} i \sin^{2} \Omega+\cos^{2} \epsilon \sin^{2} \Omega+\cos^{2} i \cos^{2} \Omega \cos^{2} \epsilon \right. \\
    & \left. + \sin^{2} i \sin^{2} \epsilon+\cos \epsilon \sin \epsilon \cos i \cos \Omega \right) .
\label{eq_PR5_final}
\end{aligned}
\end{equation}

Equation (\ref{eq_PR5_final}) also contains a drag term because the sign of Eq.~(\ref{eq_PR5_integ}) is opposite to the direction of velocity vector. $\boldsymbol{F}_{\rm{PR}, 6}$ is given as 
\begin{equation}
    \left<\frac{\mathrm{d} a}{\mathrm{d} t}\right>_{F_{\mathrm{PR}, 6}}=-\frac{B}{c|\boldsymbol{d}|^{4}} \frac{1}{2 \pi} \frac{2}{n} \int_{0}^{2 \pi} \boldsymbol{V} \cdot \boldsymbol{R} \cdot \boldsymbol{d} \cdot \hat{\boldsymbol{t}} \mathrm{d} f.
\label{eq_PR6_integ}
\end{equation}
Considering that the particles are assumed to move in circular orbit around Mars, $\boldsymbol{V} \cdot \boldsymbol{R}=0$. Hence, this term is reduced to
\begin{equation}
    \left<\frac{\mathrm{d} a}{\mathrm{d} t}\right>_{F_{\mathrm{PR}, 6}}=0.
\label{eq_PR6_final}
\end{equation}

Now, to remove the fast variable $\Omega$, we assume that the value of $\epsilon$ is small so that terms in Eq.~(\ref{eq_PR3_final}) and Eq.~(\ref{eq_PR5_final}) containing $\sin ^{2} \epsilon$ or $\sin \epsilon$ are neglected and $\cos \epsilon$ is approximated by 1 . In particular, $\epsilon$ is $25^{\circ}$ for the current Martian obliquity, then $\sin ^{2} \epsilon=0.18$ and $\cos \epsilon=0.91$. Hence,
\begin{equation}
    \left<\frac{\mathrm{d} a}{\mathrm{d} t}\right>_{F_{\mathrm{PR}}}^{\epsilon \sim 0} \approx \frac{1}{n} \frac{B\left|\boldsymbol{V}_{\mathrm{S}}\right|\left|\boldsymbol{R}_{\mathrm{S}}\right||\boldsymbol{R}|}{c|\boldsymbol{d}|^{4}} \cos i-\frac{2}{n} \frac{B|\boldsymbol{V}|}{c|\boldsymbol{d}|^{2}}\left(1+\frac{1}{4}\left(1+\cos ^{2} i\right)\right).
\label{eq_PR_final}
\end{equation}

In summary, $\boldsymbol{F}_{\rm {PR}, 1}$, $\boldsymbol{F}_{\rm {PR}, 2}$, and $\boldsymbol{F}_{\rm {PR}, 6}$ have no contribution to the secular evolution of the semi-major axis. The only positive secular perturbation on the semi-major axis is exerted by $\boldsymbol{F}_{\rm {PR}, 3}$. The secular perturbations arising from $\boldsymbol{F}_{\rm {PR}, 4}$ and $\boldsymbol{F}_{\rm {PR}, 5}$ on the semi-major axis is negative, leading to a secular decrease in the semi-major axis. $\boldsymbol{F}_{\rm {PR}, 4}$ is approximately twice as large as $\boldsymbol{F}_{\rm {PR}, 5}$. The term $\left|\boldsymbol{F}_{\rm {PR}, 3}\right| /\left|\boldsymbol{F}_{\rm {PR}, 4}\right|$ is of the magnitude of $\left(\left|\boldsymbol{V}_{\mathrm{S}}\right||\boldsymbol{R}|\right) /(|\boldsymbol{d}||\boldsymbol{V}|)$. According to \cite{Rub13}, its value is of $10^{-5}$ for Mars. Thus, their overall effects result in a decrease of the semi-major axis of the particle's motion.

We note that, with the above averaging method, the orbital elements apart from the true anomaly $f$ are assumed to be constant in one period of the particle's motion. This is an unavoidable limitation of the averaging method because the oscillating magnitude of the other four orbital elements is varying in one period of the particle's motion around Mars although the change can be small. Thus, this approximation (i.e., the averaging method) would lead to a certain deviation of the analytical results from the numerical ones. A comparison to numerical results is necessary to estimate to what degree the analytical decrement of the variation in $a$ by the PR force matches. If the analytic result matches the numerical simulation well, an accurate prediction on the particle's semi-major axis can be given. We will further discuss this point in the next section.

Using Eq.~(\ref{eq_e_Force}) and Eq.~(\ref{eq_secu_e}), the analytical description of the eccentricity evolution involves elliptic integral terms. In Eq.~(\ref{eq_e_Force}), the second term is given as
\begin{equation}
    \frac{1}{a v} r \sin f \cdot F_{\mathrm{N}}=\frac{a^{1 / 2}\left(1-e^{2}\right)^{3 / 2} \sin f \cdot F_{\mathrm{N}}}{\sqrt{G m_{\mathrm{S}}} \sqrt{\left(1+e^{2}+2 e \cos f\right)}(1+e \cos f)^{3 / 2}}.
\label{eq_Fm}
\end{equation}
Solving Eq.~(\ref{eq_Fm}) is challenging without approximation and specific expressions are not given mathematically.

To cope with the complexity in Eq.~(\ref{eq_Fm}), \cite{Mak05} introduced an analytical resolution of $\left<\frac{\mathrm{d} a}{\mathrm{d} t}\right>_{F_{\mathrm{PR}}}$ to the orbit-averaged equations of motion that are initially used in \cite{Kri96} under Martian $J_{2}$ and the SRP. Though a qualitative agreement with respect to the numerical results can be achieved for the amplitude and period of oscillations of eccentricity, only the deviation terms in the averaged eccentricity that are caused by the decay of averaged semi-major axis are considered in their work. Such treatment may leave some risk in an inaccurate description on the averaged eccentricity by ignoring those uncoupled terms. The pericenter distance $a(1-e)$ is not thoroughly expressed with the mentioned unknown variations related to the eccentricity. Thus, different from their treatment, we integrate the full dynamics of particles around Mars to better understand the long-term evolution of pericenter distance.

\section{Effects of planetary shadow}
\label{sec:sun_shadow}

The second term of Eq.~(\ref{eq_a_sunshadow_expand}) indicates the decay rate of the semi-major axis when a particle is under the shadow area due to each component $\boldsymbol{F}_{{\rm PR}, i}$ (Appendix \ref{sec:PR}), which can be calculated as
\begin{equation}
    \left<\frac{\mathrm{d} a}{\mathrm{d} t}\right>_{F_{\rm {PR}}, i, \rm shadow}=\frac{1}{2 \pi} \frac{2}{n} \int_{\psi-\phi}^{\psi+\phi} \boldsymbol{F}_{\rm {PR}, \rm i}(f) \cdot \hat{\boldsymbol{t}} \mathrm{d} f,
\label{eq_PR_inshadow}
\end{equation}
where $i=1,2, \ldots, 6$. This can be deduced from Eq.~(\ref{eq_PR_each}) that the magnitude of $\left|\boldsymbol{F}_{\rm {PR}, 3}\right|/\left|\boldsymbol{F}_{\rm {PR}, 4}\right|$ is of the order of $\left(\left|\boldsymbol{V}_{\rm S}\right||\boldsymbol{R}|\right) /\left(|\boldsymbol{V}|\left|\boldsymbol{R}_{\rm S}\right|\right)$. The one of $\left|\boldsymbol{F}_{\rm {PR}, 6}\right|/\left|\boldsymbol{F}_{\rm {PR}, 5}\right|$ is of the order of $|\boldsymbol{R}| /\left|\boldsymbol{R}_{\rm S}\right|$.

For a particle orbiting around Mars, $\left|\boldsymbol{V}_{\rm S}\right||\boldsymbol{R}| \ll|\boldsymbol{V}|\left|\boldsymbol{R}_{\mathrm{S}}\right|$ and $|\boldsymbol{R}| \ll\left|\boldsymbol{R}_{\mathrm{S}}\right|$. So, the effects of the shadowed $\boldsymbol{F}_{\rm {PR}, 3}$ and $\boldsymbol{F}_{\rm {PR}, 6}$ are neglected compared with $\boldsymbol{F}_{\rm {PR}, 4}$ and $\boldsymbol{F}_{\rm {PR}, 5}$. For $\boldsymbol{F}_{\rm {PR}, 1}$, the term $\boldsymbol{V}_{\rm S}$ in its expression ensures that its long-term effect can be eliminated by averaging upon $\cos n_{\mathrm{M}} t$ and $\sin n_{\mathrm{M}} t$ in $\boldsymbol{V}_{\mathrm{S}}$. This conclusion holds as the planetary shadow is considered. Because Mars is assumed to orbit the Sun in a circular motion, i.e., $\boldsymbol{V}_{\mathrm{S}} \cdot \boldsymbol{R}_{\mathrm{S}}=0$, the effect of the shadowed $\boldsymbol{F}_{\rm {PR}, 2}$ is eliminated. Thus, only the shadowed $\boldsymbol{F}_{\rm {PR}, 4}$ and $\boldsymbol{F}_{\rm {PR}, 5}$ need to be considered. 

Given specific orientation of the nodal line and the orientation of the planetary shadow w.r.t the $x$-axis of Mars-centered inertial frame, the last terms in $\boldsymbol{F}_{\rm {PR}, 4}$ and $\boldsymbol{F}_{\rm {PR}, 5}$ (see Eq.~(\ref{eq_a_inshadow_PR4}) and Eq.~(\ref{eq_a_inshadow_PR5})) can be further simplified. When the nodal line is $45^{\circ}$ from the $x$-axis of Mars-centered inertial frame, $2 \Omega=90^{\circ}$ and $\sin 2 \Omega$ achieves its maximum (i.e., 1). The variable $\omega+\psi$ indicates the angle between the orientation of planetary shadow (OS line in Fig.~\ref{fig_shadow_conf}(b)) and the nodal line. During the Martian orbit around the Sun, the value of $\omega+\psi$ varies and it is zero when the Sun-Mars line aligns with the nodal line. Thus, under the aforementioned assumptions, the last term in Eq.~(\ref{eq_a_inshadow_PR4}), i.e., $2 \sin 2 \Omega \cos (2 \omega+2 \psi) \sin 2 \phi$ reaches its maximum $2 \sin 2 \phi$ when $2 \Omega=90^{\circ}$ and $\omega+\psi=0$ for $\sin 2 \Omega \cos (2 \omega+2 \psi)=1$. The last term in Eq.~(\ref{eq_a_inshadow_PR5}), i.e., $\frac{1}{2}\left(1-\cos ^{2} \epsilon\right) \sin (2 \Omega+2 \omega+2 \psi) \sin 2 \phi$ reaches its maximum $\frac{1}{2}\left(1-\cos ^{2} \epsilon\right) \sin 2 \phi$ for $\sin (2 \Omega+2 \omega+2 \psi)=1$. 

Thus, for a particle moving around Mars with an arbitrary orientation of the planetary shadow and the nodal line, the minimum of Eq.~(\ref{eq_a_inshadow_PR4}) and Eq.~(\ref{eq_a_inshadow_PR5}) is the results of Eq.~(\ref{eq_a_inshadow_PR4}) and Eq.~(\ref{eq_a_inshadow_PR5}) under the aforementioned assumptions. Therefore, the minimum of $\left<\frac{\mathrm{d} a}{\mathrm{d} t}\right>_{F_{\mathrm{PR}, \rm shadow}} > 0$ is the summary of the minimum of Eq.~(\ref{eq_a_inshadow_PR4}) and Eq.~(\ref{eq_a_inshadow_PR5}) and can be given analytically as
\begin{equation}
    \left<\frac{\mathrm{d} a}{\mathrm{d} t}\right>_{F_{\mathrm{PR}, \rm shadow}} \geq-\frac{1}{2 \pi} \frac{2}{n} \frac{B|\boldsymbol{V}|}{c|\boldsymbol{d}|^{2}}\left(\frac{1}{2}\left(1+\cos ^{2} \epsilon\right) \phi+2 \phi+\frac{1}{2}\left(5-\cos ^{2} \epsilon\right) \sin 2 \phi\right).
\label{eq_PR_inshadow_minimum}
\end{equation}
\\

\section{Upper bound of pericenter distance with planetary shadow}
\label{sec:upper}

Using Eq.~(\ref{eq_a_sunshadow_lowest}), $p_{\max}$ is semi-analytically given as follows. Assuming that $a=a_{0}$ at $t=0$, the upper bound of $\left<\frac{\mathrm{d} a}{\mathrm{d} t}\right>_{\rm shadow}$ at $t=0$ in a short period $\Delta_{0}$ is estimated from Eq.~(\ref{eq_a_sunshadow_lowest}). Using $\left<\frac{\mathrm{d} a}{\mathrm{d} t}\right>_{\rm shadow}$ iteratively and starting from $a=a_{0}$ at $t=0$, the upper bound of the semi-major axis $a_{N}$ and thus $p_{\max , N}$ after $N_{\text {th}}$ step (i.e., at $t=N_{\text {th}} \Delta t_{0}$) is given
\begin{equation}
    p_{\max , N}=a_{N}=a_{N-1}+\left<\frac{\mathrm{d} a}{\mathrm{d} t}\right>_{{\rm shadow},a_{N-1}} \Delta t_{N-1},
\label{eq_p_itera}
\end{equation}
where the lowest decay rate of $\left<\frac{\mathrm{d} a}{\mathrm{d} t}\right>_{{\rm shadow},a_{i}}$ $(i=1,2, \ldots, N)$ is estimated by Eq.~(\ref{eq_a_sunshadow_lowest}). In each step, $a_{i}$ is calculated iteratively by Eq.~(\ref{eq_p_itera}). $\phi$ and $n$ in Eq.~(\ref{eq_a_sunshadow_lowest}) are related to $a_{i}$. The values of $|\boldsymbol{d}|$ and $|\boldsymbol{V}|$ are approximated as the those of a particle moving on a circular orbit around Mars with zero eccentricity guaranteed by the fact that the maximum of $p_{i}$ is achieved as $e=0$. The argument of perigee becomes zero and the ascending node can be regarded as zero under the assumption of small inclination. To conclude, if $p_{\max , i}$ is smaller than (or very close to) $R_{\mathrm{M}}$, a particle surely hits Mars.\\

\section{Semi-major Axis Evolution by Shadowed SRP and Martian $J_{2}$}
\label{sec:SRP and J2}

\begin{figure}
    \centering
    \includegraphics[width=0.9\textwidth]{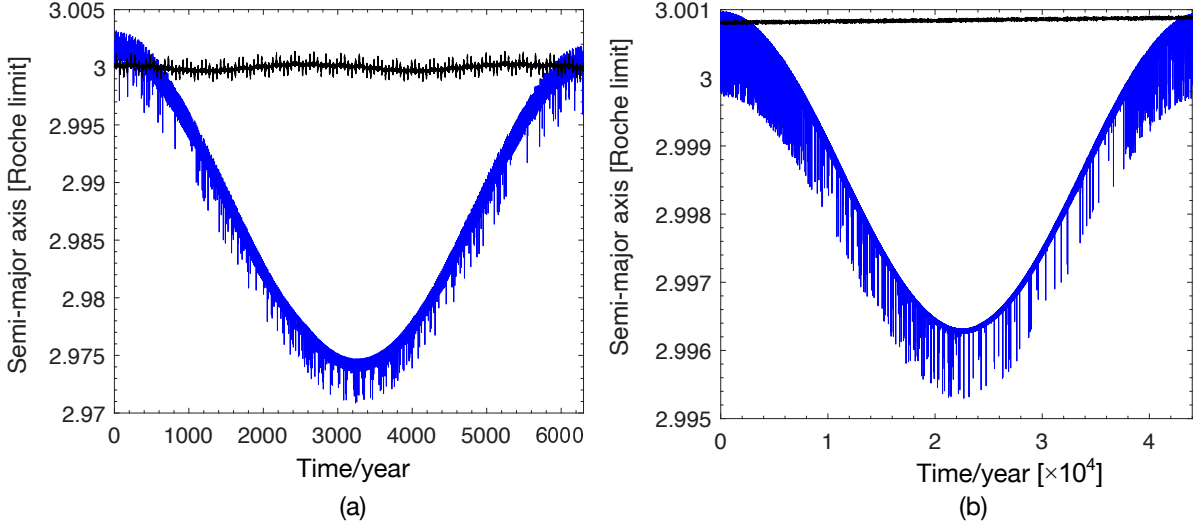}
    \caption{Evolution of semi-major axis for $r_{\rm p}=30 \mu m$ (panel (a)) and $100 \mu m$ (panel (b)) released from $a_{\rm 0}=3 a_{\rm R}$. Blue lines indicate the results under shadowed SRP alone. Black lines indicate the results under Martian $J_2$ and shadowed SRP.}
\label{fig_appendix}
\end{figure}

The shadowed SRP alone triggers a long-term periodic variation in the semi-major axis, which is first discovered by \cite{Hau13}. They proved that there does not exist an explicit solution of the secular semi-major axis that can approximate this phenomenon but their semi-analytical integrations perfectly match this periodic motion. Still, this long-term periodic variation was not mathematically proved or discussed by \cite{Hau13}. Here we try to explain why the shadowed SRP alone results in this periodic long-term behavior as follows.

The secular effect of the shadowed SRP alone on the semi-major axis can be described mathematically as
\begin{equation}
    \left<\frac{\mathrm{d} a}{\mathrm{d} t}\right>_{\rm SRP,shadow}=\frac{1}{2 \pi} \frac{2}{n} \int_{\psi+\phi}^{\psi-\phi} \boldsymbol{F}_{\rm SRP}(f) \cdot \hat{\boldsymbol{t}} \mathrm{d} f.
\label{eq_shadowSRP}
\end{equation}
The basic definition of SRP is as follows
\begin{equation}
    \boldsymbol{F}_{\rm SRP}=\frac{B}{|\boldsymbol{d}|^{3}} \boldsymbol{d}.
\label{eq_FSRP}
\end{equation}
Substituting $\boldsymbol{d}=\boldsymbol{R}+\boldsymbol{R}_{\mathrm{S}}$ and Eq.~(\ref{eq_FSRP}) to Eq.~(\ref{eq_shadowSRP}) yields
\begin{equation}
    \left<\frac{\mathrm{d} a}{\mathrm{d} t}\right>_{\rm SRP,shadow}=\frac{1}{2 \pi} \frac{2}{n} \frac{B}{|\boldsymbol{d}|^{3}} (\int_{\psi+\phi}^{\psi-\phi} \boldsymbol{R}_{\mathrm{S}} \cdot \hat{\boldsymbol{t}}\mathrm{d} f+ \int_{\psi+\phi}^{\psi-\phi} \boldsymbol{R} \cdot \hat{\boldsymbol{t}} \mathrm{d} f).
\label{eq_a_SRP_inshadow}
\end{equation}
Hence,
\begin{equation}
    \int_{\psi+\phi}^{\psi-\phi} \boldsymbol{R}_{\mathrm{S}} \cdot \hat{\boldsymbol{t}} \mathrm{d} f=\left|\boldsymbol{R}_{\mathrm{S}}\right| \cdot 2 \sin \phi \cdot\left(\cos \left(\lambda_{\mathrm{S}}-\Omega\right) \sin (\omega+\psi)-\cos i \sin \left(\lambda_{\mathrm{S}}-\Omega\right) \cos (\omega+\psi)\right).
\label{eq_Rs&t}
\end{equation}
If the values of $\omega$ and $\Omega$ are regarded as the one at the shadow entrance temporarily for the convenience of concise expression. Given a small value of inclination $i$, it can be further cast to a simpler form as
\begin{equation}
    \int_{\psi+\phi}^{\psi-\phi} \boldsymbol{R}_{\mathrm{S}} \cdot \hat{\boldsymbol{t}} \mathrm{d} f=\left|\boldsymbol{R}_{\mathrm{S}}\right| \cdot 2 \sin \phi \cdot \sin \left(\omega+\psi+\Omega-\lambda_{\mathrm{S}}\right).
\label{eq_Rs&t_i}
\end{equation}
The final expression of the second integral in Eq.~(\ref{eq_a_SRP_inshadow}) is obtained as follows
\begin{equation}
    \int_{\psi+\phi}^{\psi-\phi} \boldsymbol{R} \cdot \hat{\boldsymbol{t}} \mathrm{d} f=|\boldsymbol{R}| \cdot \frac{1}{2}\left(1-\cos ^{2} i\right)(\phi+\sin (2 \omega+\psi) \cdot \sin \phi),
\label{eq_Rs&t_final}
\end{equation}
which is neglectable compared with the first integral. The value of $\left<\frac{\mathrm{d} a}{\mathrm{d} t}\right>_{\text{SRP}}$ is mainly determined by the first integral, i.e.,
\begin{equation}
    \left<\frac{\mathrm{d} a}{\mathrm{d} t}\right>_{\text{SRP}}\simeq \frac{1}{2 \pi} \frac{2}{n} \frac{B}{|\boldsymbol{d}|^{3}}\left|\boldsymbol{R}_{\mathrm{S}}\right| \cdot 2 \sin \phi \cdot \sin \left(\omega+\psi+\Omega-\lambda_{\mathrm{S}}\right),
\label{eq_a_SRP_final}
\end{equation}
where $\sin \phi>0$. It is important to mention that $\phi=\left(E_{2}-E_{1}\right)/2$ where $E_{1}$ and $E_{2}$ denote the eccentric anomaly at shadow entrance and exit epoch. Thus, as $E_{2}=E_{1}$, a situation without planetary shadow leads to zero $\sin \phi$ and thereby zero deviation on $\left<\frac{\mathrm{d} a}{\mathrm{d} t}\right>_{\rm SRP}$.  Equation (\ref{eq_a_SRP_final}) was also derived by \cite{Hau13} (i.e., their Eq. (40)). Then, they did not discuss it mathematically, but obtained the evolution of the averaged semi-major axis by semi-analytical integration based on their Eq. (40). Here, we try to discuss why this long-term periodic motion occurs when only the shadowed SRP is considered and the dependence on its amplitude and period.

Based on Eq.~(\ref{eq_a_SRP_final}), the averaging change of semi-major axis in a certain time span $T$ can be expressed as
\begin{equation}
    \Delta a=\frac{2}{n} \frac{B}{|\boldsymbol{d}|^{3}} \sum_{i=1}^{N} K_{i} \sin \left(\omega_{i}+\psi_{i}+\Omega_{i}-\lambda_{S_{i}}\right) \Delta T_{i},
\label{eq_delt_a}
\end{equation}
where $\Delta T_{i}$ indicates the period of the particle's motion around the planet, e.g., Mars, with planetary shadow and $N$ indicates the number of such period. $X_{i}$ denotes the averaging value of parameter $X$ in the $i_{\rm th}$ period. $K_{i}=\left|\boldsymbol{R}_{S_{i}}\right| \cdot 2 \sin \phi_{i}$ can be regarded as a positive coefficient in each period. Thereby, $\Delta a$ is an accumulating sum of a series of sin function.

Hence, if $\left(\omega_{i}+\psi_{i}+\Omega_{i}-\lambda_{S_{i}}\right) \in(0, \pi), \sin \left(\omega_{i}+\psi_{i}+\Omega_{i}-\lambda_{S_{i}}\right)>0$, the sum of these series terms is increased by adding a positive term. On the contrast, if $\left(\omega_{i}+\psi_{i}+\Omega_{i}-\lambda_{S_{i}}\right) \in(-\pi, 0), \sin \left(\omega_{i}+\psi_{i}+\Omega_{i}-\lambda_{S_{i}}\right)<0$, the sum of these series terms is decreased by adding a negative term. A necessary condition of the decreasing segment in the long-term vibration is that, during a certain period of time, the accumulating effect of $\sin \left(\omega_{i}+\psi_{i}+\Omega_{i}-\lambda_{S_{i}}\right)$ is negative. For the increasing segment, the positive $\sin \left(\omega_{i}+\psi_{i}+\Omega_{i}-\lambda_{S_{i}}\right)$ is dominant.

The resultant accumulating effect is a long-term oscillation pattern with a certain period and amplitude, which is also shown by the semi-analytically integration by \cite{Hau13}. The decreasing and increasing speed of $\sin \left(\omega_{i}+\psi_{i}+\Omega_{i}-\lambda_{S_{i}}\right)$, i.e., the period of this oscillation, depend on the variation in both $\omega_{i}$ and $\Omega_{i}$. According to \cite{Kri96}, $\left<\frac{\mathrm{d} \omega}{\mathrm{d} t}\right>_{J_{2}}$ and $\left<\frac{\mathrm{d} \Omega}{\mathrm{d} t}\right>_{J_{2}}$ do not rely on the particle size while $\left<\frac{\mathrm{d} \omega}{\mathrm{d} t}\right>_{\mathrm{SRP}} \propto \frac{1}{r_{\mathrm{P}}}$ and $\left<\frac{\mathrm{d} \Omega}{\mathrm{d} t}\right>_{\mathrm{SRP}} \propto \frac{1}{r_{\mathrm{P}}}$. Under shadow condition, such proportional (positive and linear) relation turns to a positive (but not linear) correlation because the $\left<\frac{\mathrm{d} \omega}{\mathrm{d} t}\right>_{\mathrm{SRP}}$ and $\left<\frac{\mathrm{d} \Omega}{\mathrm{d} t}\right>_{\mathrm{SRP}}$ themselves are related to the distance to the Sun and the accumulating time. Explicitly expressed by Eq.~(\ref{eq_delt_a}), the decrement and increment in each period depend on the $\left|\boldsymbol{R}_{S_{i}}\right| /|\boldsymbol{d}|^{3}$, related to the averaged distance to Sun, the accumulating time $\Delta T_{i}$, related to the shadow length, and $B$, which is $\propto \frac{1}{r_{\mathrm{p}}}$. Our discussions on the dependence on the particle size agrees with the numerical results of \cite{Hau13} (see their Fig. 3). Besides the particle size, they also numerically showed the dependence on the initial semi-major axis of particle orbit.

In short, this long-term periodic pattern is caused by an accumulating effect of the shadowed SRP alone. The sum of continuous positive terms $\sin \left(\omega_{i}+\psi_{i}+\Omega_{i}-\lambda_{S_{i}}\right)>0$ leads to an increasing segment of oscillation while switching to continuous negative terms $\sin \left(\omega_{i}+\psi_{i}+\Omega_{i}-\lambda_{S_{i}}\right)<0$ leads to a decreasing segment of oscillation. The variations in $\omega_{i}$ and $\Omega_{i}$ depend on $\left<\frac{\mathrm{d} \omega}{\mathrm{d} t}\right>_{\mathrm{SRP}}$ and $\left<\frac{\mathrm{d} \Omega}{\mathrm{d} t}\right>_{\mathrm{SRP}}$.

Now we try to discuss the evolution of semi-major axis when both the shadowed SRP and the $J_{2}$ term are accounted. The semi-analytical integration of \cite{Hau13} showed that the averaged semi-major axis becomes nearly-constant (without the mentioned long-term periodic variation) but they did not give a reasonable explanation on that change.

With Martian $J_{2}$ added,
\begin{equation}
    \dot{\omega}_{i}=\dot{\omega}_{i}^{\mathrm{SRP}}+\dot{\omega}_{i}^{J_{2}}
\label{equ_domeg}
\end{equation}
and
\begin{equation}
    \dot{\Omega}_{i}=\dot{\Omega}_{i}^{\mathrm{SRP}}+\dot{\Omega}_{i}^{J_{2}}.
\label{equ_dOMEG}
\end{equation}
According to \cite{Kri96}, the following equation holds
\begin{equation}
    \dot{\Omega}_{i}^{J_{2}}+\dot{\omega}_{i}^{J_{2}}=\beta n
\label{equ_domeg&dOMEG}
\end{equation}
where $\beta=\frac{3}{2} J_{2}\left(\frac{R_{\rm {M}}}{a}\right)^{2} \frac{n}{n_{\text {M}}}$. Thus, the variations of $\Omega$ and $\omega$ are not only determined by $\left<\frac{\mathrm{d} \omega}{\mathrm{d} t}\right>_{\rm SRP}$ and $\left<\frac{\mathrm{d} \Omega}{\mathrm{d} t}\right>_{\rm SRP}$. Consequently, they result in a switch of the sign of $\sin \left(\omega_{i}+\psi_{i}+\Omega_{i}-\lambda_{S_{i}}\right)$ from positive to negative and vise versa. The previous positive term of $\sin \left(\omega_{i}+\psi_{i}+\Omega_{i}-\lambda_{S_{i}}\right)$ is no longer unchanged after adding Martian $J_{2}$. Thus, the sign of $\sin \left(\omega_{i}+\psi_{i}+\Omega_{i}-\lambda_{S_{i}}\right)$ changes. Thereby, the sum of these continuous positive or negative terms is broken, which causes that such periodic oscillation is broken and replaced by a flattened pattern.

We set $a_{0}=3a_{\rm R}$,  $e_{0}=0$, and $r_{\mathrm{p}}=30 \mu m$ and $100 \mu m$ to numerically show the aforementioned mathematical arguments. Figure \ref{fig_appendix} illustrates the time history of semi-major axis for particles with $r_{\mathrm{p}}=30 \mu m$ (panel (a)) and $100 \mu m$ (panel (b)), released from $a_{0}=3a_{\rm R}$. The blue lines in Fig.~\ref{fig_appendix} (panel (a)) and Fig.~\ref{fig_appendix} (panel (b)) indicates the results under the shadowed SRP alone. The black lines indicate the results under the shadowed SRP and Martian $J_{2}$. Both panels present the long-term vibration of the secular semi-major axis (blue lines) driven by the shadowed SRP alone. Such periodic pattern is then flattened after Martian $J_{2}$ is introduced to the simulations (see black lines).

The shadowed SRP alone produces a long-term periodic variation in the semi-major axis. With both the $J_{2}$ term and the shadowed SRP considered, this periodic variation is flattened. It suggests that the shadowed SRP alone is not neglectable. However, the shadowed SRP and the $J_{2}$ term together can be ignored, producing an approximation of the nearly constant long-term evolution in the semi-major axis.


\bibliography{Liang_Hyodo_2022_Icarus}{}
\bibliographystyle{aasjournal}



\end{document}